# Pluto's Surface Mapping using Unsupervised Learning from Near-Infrared Observations of LEISA/Ralph

A. Emran*[1], C. M. Dalle Ore[2], C. J. Ahrens[3], M. K. H. Khan[4], V. F. Chevrier[1], and D. P. Cruikshank[5]

[1] Space and Planetary Sciences, University of Arkansas, Fayetteville, AR 72701, USA.
[2] Carl Sagan Center at the SETI Institute, Mountain View, CA 94043, USA.
[3] NASA Goddard Space Flight Center, Greenbelt, MD 20771, USA.
[4] Department of Mathematical Sciences, University of Arkansas, Fayetteville, AR 72701, USA.
[5] Department of Physics, University of Central Florida, Orlando, FL 32816, USA.

**Abstract**

We map the surface of Pluto using an unsupervised machine learning technique using the near-infrared observations of the LEISA/Ralph instrument onboard NASA's New Horizons spacecraft. The principal component reduced Gaussian mixture model was implemented to investigate the geographic distribution of the surface units across the dwarf planet. We also present the likelihood of each surface unit at the image pixel level. Average I/F spectra of each unit were analyzed – in terms of the position and strengths of absorption bands of abundant volatiles such as $N_2$, $CH_4$, and CO and nonvolatile $H_2O$ – to connect the unit to surface composition, geology, and geographic location. The distribution of surface units shows a latitudinal pattern with distinct surface compositions of volatiles – consistent with the existing literature. However, previous mapping efforts were based primarily on compositional analysis using spectral indices (indicators) or implementation of complex radiative transfer models, which need (prior) expert knowledge, label data, or optical constants of representative endmembers. We prove that an application of unsupervised learning in this instance renders a satisfactory result in mapping the spatial distribution of ice compositions without any prior information or label data. Thus, such an application is specifically advantageous for a planetary surface mapping when label data are poorly constrained or completely unknown, because an understanding of surface material distribution is vital for volatile transport modeling at the planetary scale. We emphasize that the unsupervised learning used in this study has wide applicability and can be expanded to other planetary bodies of the Solar System for mapping surface material distribution.

**Unified Astronomy Thesaurus concepts:** Pluto (7777); Surface ices (2117); Astronomy data analysis (1858); Astrostatistics techniques (1886); Astroinformatics (78)

**Introduction**

Pluto, a dwarf planet and largest known trans-Neptunian object (TNO), is one of the fascinating outer solar system bodies in the Kuiper Belt. The dwarf planet exhibits abundances of volatile and nonvolatile ices on its surface with different spatial abundance and geographic distributions (Stern et al., 2015; Moore et al., 2016; Grundy et al., 2016). The presence and distribution of these various ice compositions are associated with some form of surface geological processes and seasonal interaction with the tenuous atmosphere, driven by the climate cycle of the dwarf planet (Moore et al., 2016; Schmitt et al., 2017; Binzel et al., 2017). Thus, mapping the spatial distribution of the ices and their relationship with surface geology and atmospheres is vital for a better understanding of the volatile transport models on the dwarf planet (Bertrand and Forget 2016). Accordingly, over the past decades, the spatial distribution of surface materials has been observed from the spectra at visible (VIS) and near-infrared (NIR) wavelengths from both ground-based telescopes and New Horizons fly-by to analyze the geographic distribution of ices on Pluto (see Cruikshank et al., 2015, 2019, and 2021 for a comprehensive review).

In the conditions on Pluto's surface, nitrogen ($N_2$), methane ($CH_4$), and carbon monoxide (CO) ices are the most abundant volatile materials (Stern et al., 2018). Among these volatiles, $CH_4$ ice has several strong absorptions bands at NIR wavelengths, such as 1.30 – 1.43 μm (region 1), 1.59 – 1.83 μm (region 2), 1.90 – 2.0 μm (region 3), and 2.09 – 2.48 μm (region 4) which have been utilized to characterize $CH_4$ ice on Pluto by previous studies (see Schmitt et al., 2017 and Scipioni et al., 2021). These characteristic absorption bands of $CH_4$ are also covered by the near-infrared spectrometer instruments onboard NASA's New Horizons spacecraft. $N_2$ ice can be detected by the presence of a 2.15-μm absorption band and CO by the 2.35-and 1.58-μm absorption bands, which have also been observed by ground-based facilities (Owen et al., 1993). Besides the volatile materials of $CH_4$, $N_2$, and CO, Pluto's surface hosts photochemically formed carbonaceous organic disordered polymeric solid called tholins and a trace amount of ethane ($C_2H_6$) and water ($H_2O$) ice (Merlin, 2015; Cook et al., 2019). $H_2O$ ice acts as nonvolatile ice on Pluto's surface and the geology of rugged terrain and undulating morphology of the dwarf planet indicate that $H_2O$ ice forms the supporting bedrock for the surface volatiles (Stern et al., 2015; Spencer et al., 2020). The equatorial-colored area of Pluto represents the residue of deposited tholins which are assumed to form due to the ultra-violet (UV) or charged-particle irradiation and energetic processing of (both likely atmospheric or surface processes) methane and nitrogen (Cruikshank et al., 2005).

Before NASA's New Horizons spacecraft's flyby in 2015, Pluto was studied from ground-based observation or space telescope such as Hubble Space Telescope (HST). For instance, methane was first identified on the surface of Pluto from ground-based observations (Cruikshank et al., 1976), followed by the identification of nitrogen and carbon monoxide ice (Owen et al., 1993). However, owing to the lack of the resolution needed, a detailed disk resolved surface mapping was not possible with these ground and space-based observations. Thus, constraining the latitudinal distribution of surface ices was not possible due to the dearth of resolution needed to resolve the surface. However, the zonal (longitudinal) distribution of ices over time from ground-based telescopes has been reported by Grundy et al. (2013, 2014). NASA's New Horizons spacecraft returned an unprecedented amount of data to Earth for analysis that paved the way to study the surface of Pluto (and its satellites) in finer detail. The initial results of the flyby mission can be found in Stern et al. (2015) and Grundy et al. (2016), and a comprehensive review can also be found in Stern et al. (2018) and Stern et al. (2021).

Efforts have been put forward to map the surface composition and materials on Pluto from the NIR hyperspectral spectra from the Linear Etalon Imaging Spectral Array (LEISA) instrument, a part of the Ralph instrument package (Reuter et al. 2008) onboard New Horizons. For instance, Grundy et al. (2016) used band depths measurement to map the surface distribution of $CH_4$, $N_2$, and CO on Pluto's encounter hemisphere. Schmitt et al (2017) used principal component analysis (PCA), spectral indicators, and correlation plots to study the distribution and physical state of ices, both volatile and nonvolatile, on Pluto's illuminated disk. Protopapa et al. (2017) used radiative transfer modeling (Hapke, 1993) on LEISA spectra to derive the spatial distribution and grain sizes of volatile and nonvolatile ices on Pluto's surface. Recently, Gabasova et al. (2021) used the intensity-based registration technique to the lower resolution data to make a global qualitative compositional map for surface ices. However, existing mapping efforts on the spatial distribution of ices are mostly accomplished from the compositional analysis using spectral indices (e.g., band depth) or an application of complex radiative transfer modeling (RTMs). An implication of these techniques requires reliable label data, spectral endmembers, or prior knowledge about the spectral characteristics of the representative surface composition.

The convenience of a machine learning technique lies in the potential to analyze large dimensional datasets more efficiently and effectively. Unsupervised learning, such as a clustering analysis (Marzo et al., 2006, 2008, 2009), is one of the widely used methods adopted to map the surface of planetary bodies. The unique advantage of unsupervised learning is that it does not need any label data or prior information. The unsupervised method utilizes the inherent characteristics of the data itself to recognize unseen but meaningful information in the data. Unsupervised spectral K-means clustering has successfully been used to characterize the surface of Mars (Marzo et al., 2006), asteroids Bennu (Rizos et al., 2019) and Ceres (Rizos et al., 2021), Saturn's moon Iapetus (Pinilla-Alonso et al., 2011), Pluto (Dalle Ore et al., 2019) and its largest satellite Charon (Dalle Ore et al., 2018). However, even though having the merit of fast computation and satisfactory results in many instances, the K-means clustering has its inherent limitations. The K-means clustering uses a rigid segmentation algorithm which may not effectively be used with many real-world datasets. Moreover, the K-means algorithm may not capture the inherent heterogeneity of the datasets and does not work well if the data have a complex non-linear pattern. See Patel and Kushwaha (2020) for an assessment of the performance of K-means clustering.

As an alternative unsupervised clustering and classification scheme, the Gaussian mixture model (GMM; Berge and Schistad Solberg, 2006; Li et al., 2014) has been used for hyperspectral images like the LEISA data. GMM uses a probabilistic approach which assumes each data point comes from a mixture of a finite normal distribution with unknown parameters (Patel and Kushwaha, 2020). GMM can uncover complex patterns, trace inherent heterogeneity, and group the data into cohesive components such that the clusters are a very close representation of the real world (Patel and Kushwaha, 2020). Moreover, for overlapping clusters, GMM performs considerably better than the K-means algorithm because K-means split the data into $k$ non-overlapping groups while the GMM can infer probabilities for the overlapping clusters.

However, the implementation of GMM to higher dimensionality (i.e., hyperspectral) data constrains its wide applicability due to the impractically large size of the parameter space (Li et al., 2014) and to the component distribution, formalized as a probability density function in a mixture model (Alqahtani and Kalantan, 2020). Moreover, working with big and higher-dimensional data is challenging, computationally expensive, and time-consuming. Thus, efficient dimensional reduction methods such as principal component analysis (PCA; Pearson, 1901) are widely used to reduce data dimension, keeping much of the possible variance in the original data, which can further be used for classification or clustering schemes. Of late, similar implementation of principal component reduced Gaussian mixture models (hereafter we refer to as PC-GMM), an unsupervised machine learning approach, has demonstrated to be satisfactory in many scientific fields (Alqahtani and Kalantan, 2020; Hertrich et al., 2022).

This study implements the PC reduced GMM to map the surface distribution of ices. The basic framework of PC-GMM used in this study is that we first apply PCA to reduce the higher dimension LEISA data to a lower dimension and then implement the GMM to that principal component reduced data. We provide the likelihood of the surface units at pixel scales, which gives an insight into a deeper understanding of the distribution of ices on Pluto's surface. We then extract I/F spectra of the surface units and analyze them in terms of the position and strengths of absorption bands of abundant volatiles such as $N_2$, $CH_4$, and CO and nonvolatile $H_2O$. An interpretation of the connection of the clusters to surface composition, geology, and location (latitudes and longitudes) is provided subsequently. We also compare the result of surface mapping from the PC-GMM to the existing surface compositional mapping results.

**Observations**

We use spatially resolved hyperspectral scenes from the LEISA instrument (Reuter et al., 2008) that cover the entire disk of Pluto. LEISA is a wedged etalon infrared spectral imager with a 256 x 256-pixel detector array, scans at 0.9º × 0.9º FOV, and operates in a push-broom mode across the target using the motion of the spacecraft. The instrument filter has two segments: one with wavelength range of 1.25 - 2.5 µm and another of 2.1 - 2.25 µm at spectral resolving powers (wavelength / Δ wavelength) $\lambda/\Delta\lambda$ = 240 and 560, respectively. We use the lower resolution wavelength segment (1.25 - 2.5 µm with the $\lambda/\Delta\lambda$ = 240), which has been used to map the distribution of ices of $CH_4$, $N_2$, CO, and $H_2O$ over the surface of the dwarf planet and $H_2O$ frost on Pluto's largest satellite Charon. The two LEISA scenes, originally at a spatial resolution of 6 and 7 km/pixels, used in this study were taken at ~100,000 km from the surface during the New Horizon's Pluto encounter period in July 2015. The details of the datasets used are given in Table 1.

**Table 1.** Details of the two LEISA scenes used in this study.

| MET | Scan name | UT date and time | Range (km) | Sub S/C Lon (°) | Sub S/C Lat (°) |
|---|---|---|---|---|---|
| 0299172014 | P_LEISA_Alice_2a | 2015-07-14 09:33:05 | 112742 | 158.62 | 38.52 |
| 0299172889 | P_LEISA_Alice_2b | 2015-07-14 09:48:16 | 100297 | 158.81 | 37.91 |

The two image scenes were processed and calibrated with the mission data pipeline processing routine including bad pixel masking, background noise cleaning, flat fielding, and conversion of DN to radiance factor (*RADF*; commonly known as I/F). The images were then projected to a common orthographic viewing geometry appropriate to the mid-time between them using the United States Geological Survey's (USGS) Integrated Software for Imagers and Spectrometers (ISISv3) software package. For details on the LEISA data processing and reduction, please see Schmitt et al. (2017), Protopapa et al. (2017), and Cook et al. (2019).

We use the derived LEISA data (I/F) products generated by the New Horizons Pluto encounter surface compositional science theme team (Stern, 2018) available at NASA PDS: Small Bodies Node[2]. The dataset consists of the spatial-spectral I/F cubes at the spatial dimension of 800 x 800 pixels and the spectral dimension of 256 wavelength bands. Note that the derived data products were reprojected at a higher spatial resolution (using the nearest neighbor resampling algorithm) than the native LEISA data to minimize the loss of spatial information during projecting the scenes in orthographic viewing geometry (Schmitt et al., 2017). Thus, the derived product we used in this study has about 2 – 3 times greater spatial resolution than the original. The lower spectral resolution of the 1.25 - 2.5-µm wavelength segment used in the study consists of 0 - 196 channels (# 200 – 255 channels that cover wavelengths between 2.10 and 2.25 µm at higher spectral resolution). Note that some of the spectral channels (e.g., # 1, 2, 198) were excluded from the original I/F cube because of a strong photometric calibration caveat at these spectral bands (Schmitt et al., 2017). The two LEISA scenes used in this study overlap and cross through Sputnik

---

[2] https://pds-smallbodies.astro.umd.edu/

Planitia (SP) and we cut the image cubes and joined them together (Fig. 1) in a similar fashion to that by Schmitt et al. (2017).

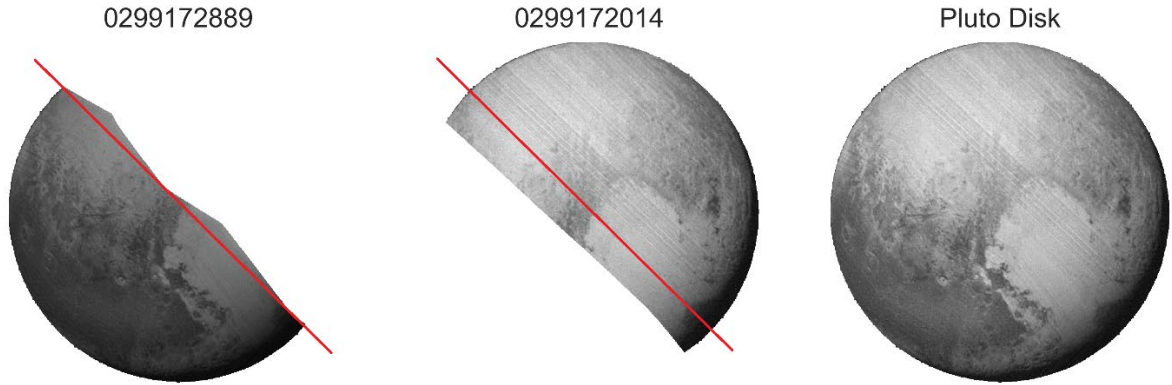

**Fig. 1:** Images of two LEISA cubes at a band at 2 µm and mosaics covering the entire Pluto disk, adopted from Schmitt et al. (2017). The image cubes are projected to a common orthographic viewing geometry appropriate to mid-time between scenes. The red line represents the overlapping line where the image cubes were cut and joined together side by side.

**Methodology**

The radiance I/F data were converted to reflectance factor (*REFF*) by dividing the I/F cube by the cosine of the incident angle as defined by Hapke (2012) following the equation for Lambertian photometry (isotropic scattering from the surface). This correction minimizes the effect of solar irradiation on the reflected radiance and represents the intrinsic photometric properties of surface materials, and has been successfully implemented on Pluto for the same LEISA image datasets (Schmitt et al., 2017). The I/F data cubes were further calibrated by the scaling factor of LEISA radiometric calibration of 0.74±0.05 (Protopapa et al., 2020). We also excluded the pixels that correspond to an incident angle of greater than 85º (and belong to a strongly shadowed area; see the Pluto disk in Fig. 1) since such weakly illuminated areas result in a noisier reflected signal and strongly affect the photometric properties of *REFF* (Schmitt et al., 2017). Very few scattered pixels at some spectral bands (~ 0.03 % of the entire observations) show no data values inside Pluto's illuminated disk and they were filled in using the linear interpolation technique for each pixel location. Note that we did not interpolate values for all spectral bands at a pixel based on the values of the nearest pixels, rather we fill the missing spectral band values based on the values of other spectral bands at that particular pixel. We emphasize that this gap-filling of some spectral bands at a few scattered pixels does not change the original data characteristics.

**Principal component reduced Gaussian mixture model (PC-GMM)**

With the implementation of PC-GMM, we reduce the higher dimension LESIA data to a lower dimensionality dataset (of an order of tens) using a principal component analysis. For convenience, we present the conceptual frameworks of the mathematical formulation of PCA. The original input dataset is considered to be a matrix of *Y*. In what follows, $Y \in \mathbb{R}^{n \times m}$ where $Y_i = (Y_{i1}, Y_{i2}, \ldots, Y_{im})^T$ for $i = 1, 2, \ldots, n$, and $m$ is the data dimension. If the rank of the matrix is $r$ where $r \leq min(n, m)$, then the principal components (Abdi and Williams, 2010) of *Y* can be computed using the singular value decomposition given by:

$$Y = U\Sigma W^T \quad (1)$$

where $\Sigma$ is the $r \times r$ diagonal matrix whose diagonal entries are the nonzero singular values $\{\sigma_l: l = 1, 2, \ldots, r\}$ with $\sigma_1 \geq \sigma_2 \geq \ldots \geq \sigma_r$, $U$ is the $n \times r$ matrix of left singular vectors, and $W$ is the $m \times r$ matrix of right singular vectors. Thus, the full principal components decomposition of *Y* can be written as:

$$P = YW = U\Sigma W^T W = U\Sigma \quad (2)$$

We use the scikit-learn (Pedregosa et al., 2011) python module to implement the PCA on the LEISA cube data. To this end, we first standardize the *REFF* data, since standardization is recommended before the implementation of PCA (e.g., Jolliffe and Cadima, 2016; Gewers et al., 2021). The standardization approach used here involves removing the mean and scaling the data to unit variance. We find that the axes of principal components (PC) up to the first four pc-axes encompass substantial surface information of the Pluto disk (Fig. 2). The pc-axes look largely similar to that of Schmitt et al. (2017). From pc#5 onward the pc-axes contain "noisy" data and, therefore, were not considered in the further analysis because their inclusion may mislead the further statistical analysis. The assignment of "noisy" data was accomplished subjectively by visually analyzing (qualitative) if the resultant pc-axes can represent the appearance of the underlying geomorphology of Pluto (visible structure/feature) seen in the basemap. We plot the variance for the pc-axes showing that with the first four pc-axes it covers 92.93% of the total variance of the data and incorporate a scree plot in determining the optimum number of pc-axes (see Fig. A1 in the appendix). From pc#4 onward the cumulative explained variance shows a fairly gentle increase –no substantial information will be gained if a higher number of components is included in the analysis. We consider the report of cumulative explained variance as a function of pc-axes, scree plot, and visual inspection of the pc-axis in choosing the number of pc-axes to use in this study.

Once the principal components are calculated, we implement the multivariate Gaussian mixture model to the first four pc-axes. For convenience, we present the conceptual framework for the mathematical formulation of multivariate GMM. The pc-axes are considered as a matrix of $X$. For what follows, $X \in \mathbb{R}^{n \times d}$ and $X_i = (X_{i1}, X_{i2}, \ldots, X_{id})^T$ for $i = 1, 2, \ldots, n$, and $d$ is the number of pc-axes. The multivariate Gaussian mixture distribution of $n$ independent observations can be written as:

$$X_i | z_i = j \sim N_d(X_i | \mu_j, \Sigma_j) \qquad (3)$$

where $z_i \in \{1, 2, \ldots, k\}$ is the latent variable representing the mixture component for $X_i$, $k$ is the total number of mixture components, $p(z_i = j) = \pi_j$ with $\sum_{j=1}^{k} \pi_j = 1$, $\mu_j$ and $\Sigma_j$ are the mean vector of length $d$ and the $d \times d$ covariance matrix of the $j$-th component, respectively, and $N_d$ is the $d$-dimensional Gaussian distribution. The marginal distribution of $X_i$ integrating out of $z_i$ can be written as:

$$p(X_i) = \sum_{j=1}^{k} p(X_i, z_i = j) = \sum_{j=1}^{k} p(z_i = j) p(X_i | z_i = j) = \sum_{j=1}^{k} \pi_j N_d(X_i | \mu_j, \Sigma_j)$$
$$(4)$$

which is the *k*-component multivariate Gaussian mixture model (e.g., Bishop and Nasrabadi 2006) with the mixture weights $\{\pi_j\}_{j=1}^{k}$.

If the parameter to estimate is $\theta = \{\mu_1, \mu_2, \ldots, \mu_k, \Sigma_1, \Sigma_2, \ldots, \Sigma_k, \pi_1, \pi_2, \ldots, \pi_k\}$, then the log-likelihood function can be written as:

$$l(\theta) = \sum_{i=1}^{n} \log\left(\sum_{j=1}^{k} \pi_j N_d(X_i|\mu_j, \Sigma_j)\right) \quad (5)$$

We use the expectation-maximization (EM; Dempster et al., 1977) algorithm to calculate the maximum likelihood estimator of the parameter $\theta$. For convenience, we present the detailed EM algorithm in Appendix (Section A.2).

We implement the *Gaussian Mixture Model* python module available in the scikit-learn (Pedregosa et al., 2011). The selection of the optimal number of gaussian components (GC) or clusters is one of the critical considerations for the implementation of GMM. Choosing the optimal number of clusters can be accomplished by evaluating the Akaike information criterion (AIC) and Bayesian information criterion (BIC) values for the applied models. In machine learning, BIC and AIC diagnoses are also treated as internal cluster validation measures/ indicators in evaluating the quality of the modeled clustering structure. Internal cluster validation measures (e.g., BIC) have been instigated in astrophysical data problems for the implementation of a probabilistic gaussian mixture model (e.g., de Souza et al., 2017). In this study, we utilize both BIC and AIC measures in determining the optimal number of clusters to use for GMM on the pc-axes (and as an indicator of internal cluster validation).

The AIC provides an estimation of the prediction error by an implemented model and thus indicates the relative goodness of the model in fitting the data upon which the model was built. The calculation of AIC is based on maximum likelihood criteria (Akaike, 1974) given as:

$$AIC = -2\log(\hat{L}) + 2D \quad (10)$$

where $\hat{L}$ is the maximized value of the likelihood function and $D$ is the total number of (independently adjusted) parameters. The concept of BIC is closely aligned to AIC since the BIC is also based on the maximum likelihood function and calculated as (Schwarz, 1978):

$$BIC = -2\log(\hat{L}) + \log(N)D \quad (11)$$

where $N$ is the total number of observations.

Both AIC and BIC penalize the model if it incorporates overfitting. However, the penalty is more extreme for BIC as it heavily penalizes the model complexity (Bishop and Nasrabadi, 2006). Typically, the best model minimizes AIC values; a lower AIC value corresponds to a better fit (Liddle, 2007). Conversely, the first local minimum of BIC value is considered the optimal number of clusters (Dasgupta and Raftery, 1998; Fraley and Raftery, 1998). Note that BIC values are also presented with an opposite sign by different studies (Fraley and Raftery, 1998), in which case the optimal number of clusters will be at the first local maximum. We choose the optimal number of clusters considering the AIC and BIC values – a lower AIC value and first local minimum, respectively. Accordingly, the optimal number of the cluster in this instance was set at 8 (which agrees with both the AIC and BIC criteria). A plot of AIC and BIC values at different numbers of clusters is given in Appendix (Fig. A2).

In unsupervised learning, a goodness-of-fit diagnostic is required to validate the model further in simulating the original data. We provide an assessment of the model fit for the observed data in hand and its capabilities in predicting future datasets. To that end, we first simulate the input data (synthetically reproduces the original data structure) based on our implemented GMM model parameters for different gaussian component solutions i.e., numbers of clusters. In this instance, we have observed data from the first 4 pc-axes used as the input data for GMM, and the predicted data were simulated from the resulting GMM model parameters. Then, we calculate the density for both observed and predicted data from multivariate kernel density estimates using the *kde* module of R in a Python environment. Finally, we compare the estimated kernel densities for observed data against the predicted data by utilizing a linear fit for different GC solutions (see Fig. A3 in Appendix). A similar goodness-of-fit diagnostics for the probabilistic GMM has successfully been implemented in an astronomical problem (de Souza et al., 2017). We calculate the coefficient of determination ($R^2$) between the observed and predicted kernel density for 5 – 12 GC solutions (Fig. A3). The $R^2$ values for different GC solutions suggest that 8 GC solution renders the highest value of 0.96 – which suggests a good fit for the model.

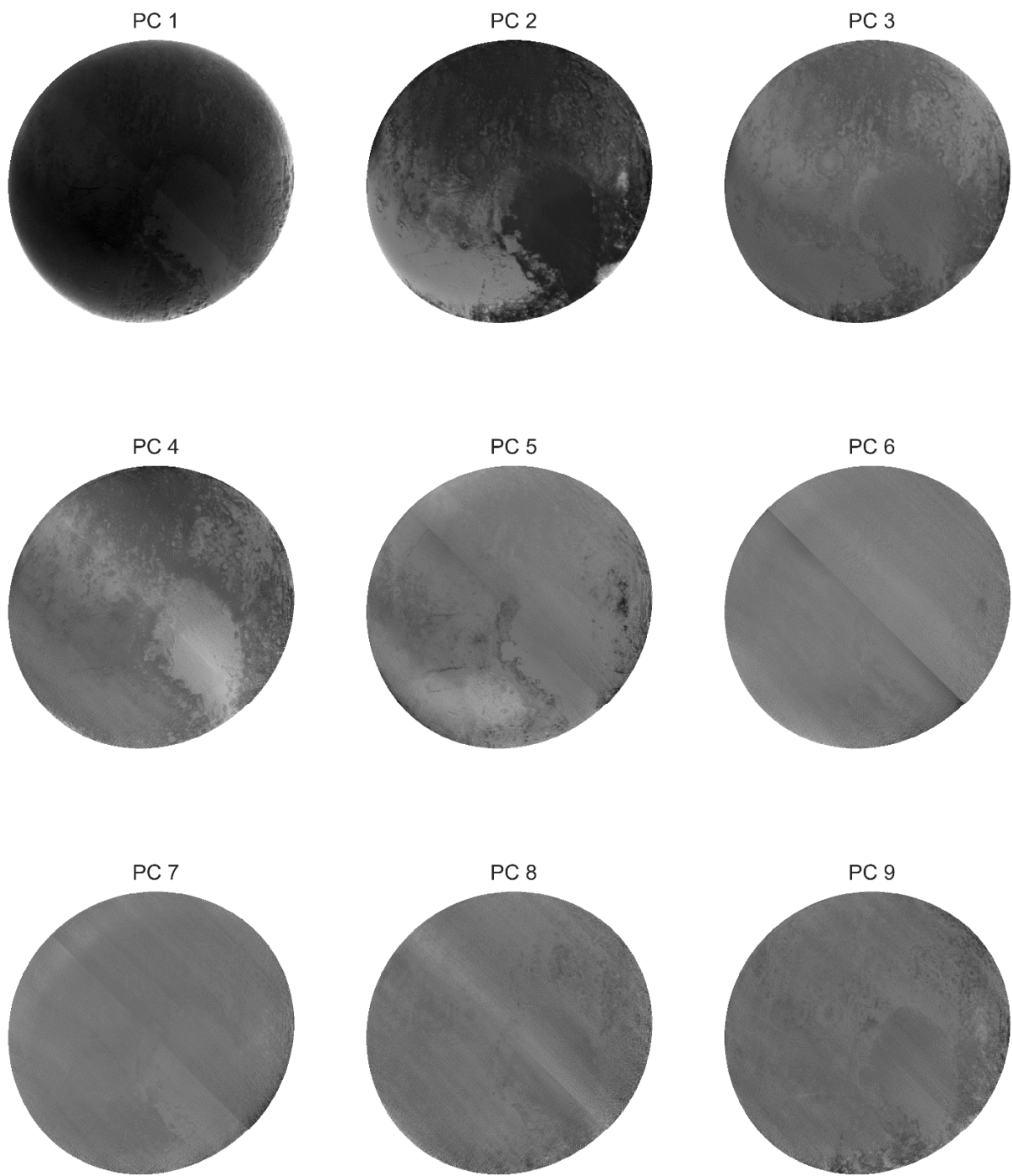

**Fig. 2:** The first nine principal components from LEISA *REFF* data cubes. Up to four pc-axes, the disks show substantial surface information (covering 92.93% total variance of the data). The pc#5 and onward components show "noisy" data therefore the pc-axes #5 and higher were not considered in the analysis. Note: the assignment of "noisy" data was accomplished subjectively by visually analyzing if pc-axes can represent the underlying geomorphology of Pluto.

**Results**

As suggested by the AIC and BIC values, we classify the entire Pluto's surface into eight representative surface units (Fig. 3). For the convenience of interpretation of the surface unit distribution, we hereafter call the surface unit by # followed by the assigned surface unit number. The surface units are distributed in different geographic regions (with different latitudinal and longitudinal extent) across the dwarf planet. Moreover, the distribution of the surface units coincides with the major geologic features e.g., basin, crater, montes, terra, etc. Note that some of the names of features and regions on Pluto used in this study are informal, while other names have been formally adopted by the Nomenclature Committee of the International Astronomical Union. We label the major geological features on the reference map in Fig. 3a, however, to reference local scale geologic features, a basemap showing the names of all geologic provinces and features on Pluto was added in Appendix (Fig. A4).

The north polar region around Lowell Regio is a distinct surface unit (#2) that extends up to Voyager Terra to the south. Roughly, the north polar unit runs from the pole to ~60º N. Likewise, the equatorial Cthulhu Macula (or simply Cthulhu) – located in the western part of SP and spanning between ~ 15º S to 12º N – exhibits a separate surface unit (#7). The surface unit #6 covers a large part of Sputnik Planitia and a portion of the northern mid-latitudes at Venera Terra (between ~ 45 to 60º N). However, the surface unit in the northern mid-latitudes at Pioneer Terra (below the Lowell Regio) belongs to #8 – extending to the latitudes of ~ 30 to 70º N with varying widths at different longitudes. Tombaugh Regio – located in the south and eastern part of SP (between ~ 15º S to 30º N latitudes) – indicates a distinct surface unit (#5). The eastern and southern limbs of the Pluto disk, including the Tartarus Dorsa region, belong to surface unit #1. This unit follows a narrow strip along the limb of Pluto's illuminated disk during the New Horizon flyby in 2015.

The broader region between the Cthulhu Macula (stretching from the latitude of Elliot crater and Virgil Fossae) and the latitude of the Burney crater shows a distinct surface unit (#4). This unit follows a latitudinal pattern and stretches between ~ 10 to 35º N. The mountains at the western fringe of the SP such as Baret Montes, Al-Idrisi Montes, Hillary Montes, etc belong to #4. Note that a narrow strip of #8 is placed (also seen in the base of the Burney crater) in between the surface units of #4 and #6. Kiladze crater, Krun Macula, inner rims of Oort crater, and the parallel stripe region following the south of the Cthulhu Macula belong to surface unit #3. The mountains (e.g.,

Pigafetta Montes) and small depressions (probably small craters) in the Cthulhu Macula and some patchy spots in Tombaugh Regio (smaller fraction) are also part of #3.

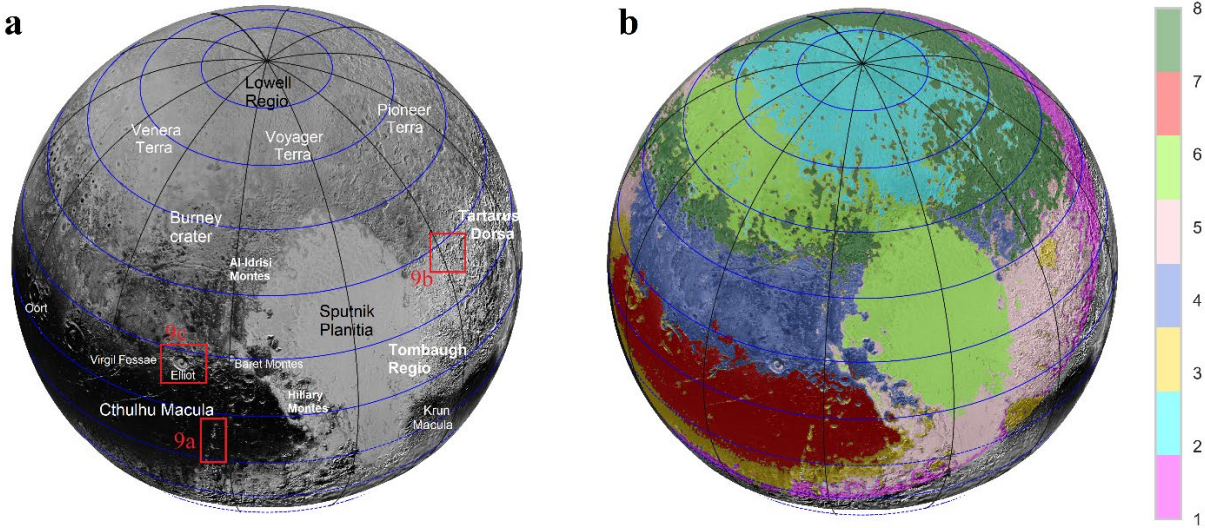

**Fig. 3:** Long Range Reconnaissance Imager (LORRI) higher resolution (panchromatic) basemap projected to LEISA scenes (a). The LORRI base map is labeled with latitude (blue) and longitude (black) grid and formal and informal names of major geomorphic features. The red rectangles represent the reference locations in Fig. 9. Eight surface units were found on Pluto using PC-GMM (b). The surface units follow the latitudinal pattern and are consistent with broad geographic regions and geologic features. The north pole at Lowell Regio, Cthulhu Macula, Tombaugh Regio, Sputnik Planitia and Venera Terra, and Pioneer Terra region show distinct surface units. The values in the legend indicate the assigned surface unit number. Note: the lower right region belongs to strongly shadowed areas and, thus, the pixels at these locations were excluded from the map (see the text for a detailed explanation).

Because the GMM employs probability distribution, we provide the likelihood of each surface unit at the LEISA pixel level (Fig. 4). While predicting the probability, the GMM algorithm treats each surface unit as a Gaussian component and infers the likelihood based on the estimated component density for every pixel. The probability subplots indicate that most of the surface units are distributed in a spatially distinct location with minimal disjointing and scattering, but with some exceptions.

The probability scale ranges from 0 to 1, where higher values indicate a higher probability for the corresponding surface unit. For pixels that are close to multiple cluster centroids, the GMM algorithm infers the probabilities of the pixels for these multiple clusters. Thus, the probability subplots in Fig. 4 indicate whether a pixel does purely belong to a particular surface unit or shares characteristics of multiple units. Most of the surface units are dominated by pixels with higher probability values (close to 1) – indicating the dominance of mostly pure pixels at each unit. This is further supported by the boxplot of the probability distributions extracted from the pixels corresponding to each class (Fig. A5 in the Appendix). As seen in the probability boxplot, the median probability of every class is more than 0.95 – confirming that each class host pixels with higher probability values.

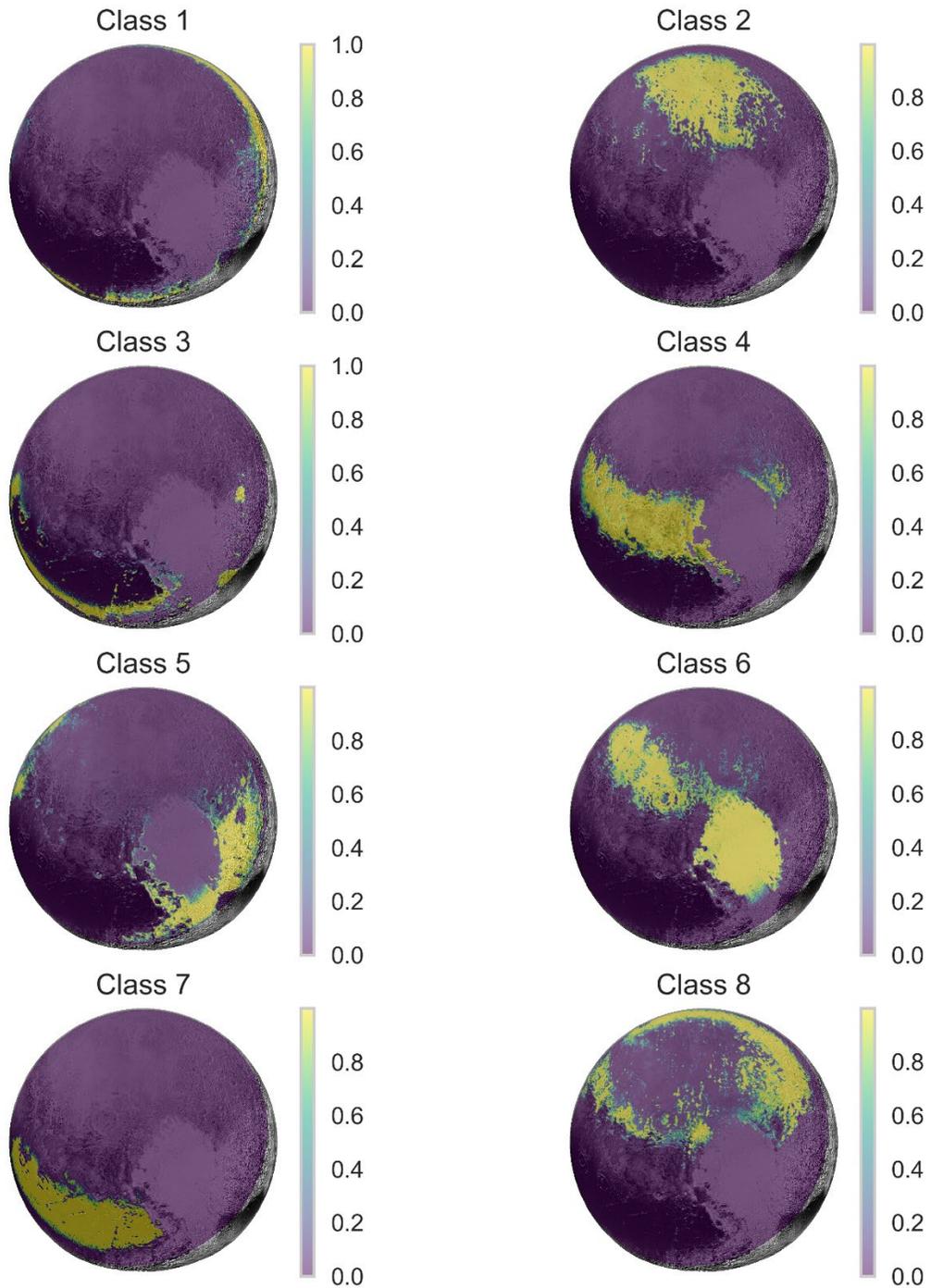

**Fig. 4:** The probability plot of each surface unit at LEISA image pixel level using PC-GMM. The subplots are superposed on the LORRI base map. The probability scale ranges from 0 to 1, where higher values indicate a higher probability of the corresponding unit. We also refer the readers to Fig. A5 in the Appendix, which shows that most of the surface units are dominated by pixels with higher probability values (close to 1).

Note that the resultant surface units using unsupervised learning are agnostic of the physical and geological meaning of the different terrains. Thus, a connection of the surface units to the chemical composition of the corresponding unit is necessary for the geologic explanation of the classification. Accordingly, we retrieve the mean and 1σ standard deviation of the I/F spectra from all the pixels that fall into each surface unit (Fig. 5). For the convenience of interpretation, the spectral subplot for the corresponding surface unit is labeled as C followed by the assigned unit number. For instance, the I/F spectra of surface unit #6 is labeled as C6. Note that these signs (C and #) are used interchangeably to denote the surface unit and corresponding spectra later in the paper for simplicity.

The standard deviation (gray shade) of the I/F spectra shows varying degrees of closeness to the mean spectra (red) at different wavelengths (see Fig. 5). The mean I/F spectra indicate the presence or absence of absorption bands of the abundant volatiles of $CH_4$, $N_2$, and CO. However, compared to $N_2$ and CO ices, $CH_4$ has very strong absorption bands at the wavelengths (1.2 to 2.5 µm) covered by the LEISA instrument. Thus, $CH_4$ absorption bands are much more readily identifiable than the other volatile ices from the mean I/F spectra of the surface unit (see Table 3 which lists important bands for different ice). All the classes, except the Cthulhu Macula (C7), exhibit a clear presence of $CH_4$ ice absorption bands with varying degrees of strength in the absorption bands. This is evidence of the widespread $CH_4$ ice distribution on Pluto's surface. The north polar unit (around Lowell Regio) – stretching from ~60° N to the north pole – shows a zone with a higher concentration of $CH_4$ ice (Grundy et al., 2016; Earle et al., 2018) as evidenced by the strong absorption features in the I/F spectra. More details of the spectral interpretations of all surface units are given in the later section of the paper.

The I/F spectra for the broad surface unit of Sputnik Planitia and Venera Terra (#6) exhibit the presence of $N_2$ and CO absorption bands at 2.15 µm and 1.58 µm, respectively (see C6 in Fig. 5). These bands are subtle at the scale shown, but are statistically reliable (because the absorption bands plotted in the figure are accompanied by a 1-σ error bar). We did not see the sharp 2.35-µm absorption band for CO at the LEISA spectral resolution since the absorption band is blended between two broad $CH_4$ absorption bands (Schmitt et al., 2017). Since both $N_2$ and CO absorption bands are much weaker compared to the $CH_4$ bands, a finer difference in the spatial distribution of $N_2$ and CO abundance may not be sharply evident when applying an unsupervised classification scheme to the entire Pluto disk data. Thus, we re-classify the N$_2$ and CO-rich surface unit (#6) into multiple subunits to probe the finer difference in the distribution of N$_2$ and CO ices using a post-classification scheme. During re-classification, we treated the surface unit (#6) independently and did not increase the original gaussian components.

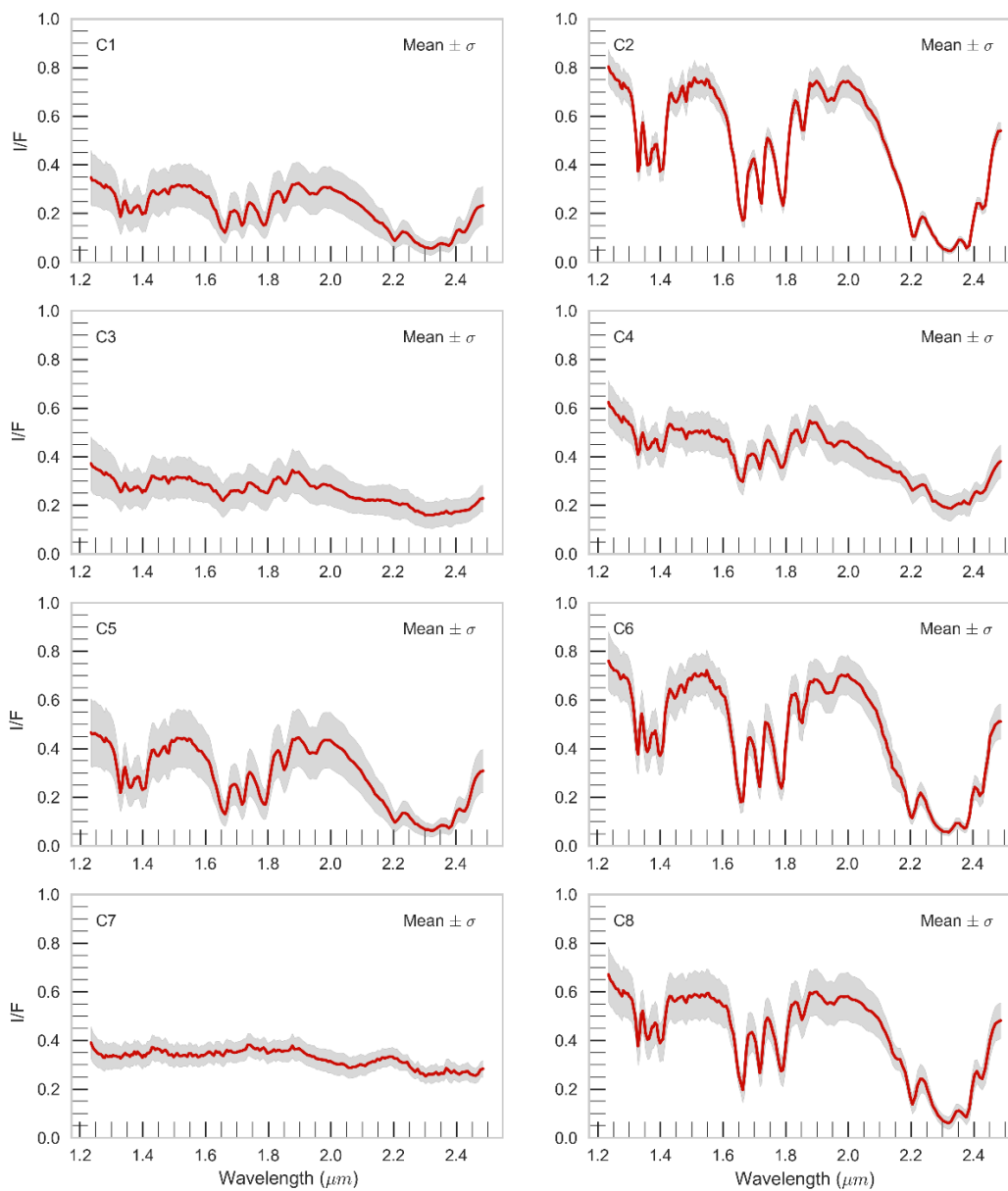

**Fig. 5:** The mean (red line) ± 1σ standard deviation (gray shade) I/F spectra of each surface unit on Pluto using PC-GMM. The spectra of the surface units in the subplots are labeled as C followed by the corresponding assigned surface unit number. The standard deviation of the I/F spectra shows varying degrees of closeness to the mean spectra at different wavelengths. All the classes (except the C7) exhibit a clear presence of CH$_4$ ice absorption bands with varying degrees of strength.

H$_2$O ice on the dwarf planet can be detected using the absorption bands at 1.65 and 2.0 µm, which are an indicator of the crystalline ice phase (Cook et al., 2019). On Pluto's surface, the volatile ices of N$_2$, CO, and CH$_4$ mostly hide the H$_2$O ice bedrock from being identified in the LEISA wavelengths (Cruikshank et al., 2021). Moreover, the spectral signature of H$_2$O on Pluto on a global scale is indiscernible because the broad spectral bands of H$_2$O are much shallower and are mostly overlain by strong CH$_4$ bands (Cruikshank et al., 2021). However, a relatively higher abundance of H$_2$O ice has been identified in different isolated areas (in smaller areal fractions) including the Kiladze crater and Krun Macula (Cook et al., 2019). These water-ice-abundant areas correspond to our surface unit #3 –which shows the presence of an absorption band at 1.65 µm in the I/F spectra (Fig. 5).

Though the spectra for C3 (see Fig. 5) do not show a very strong 2.0-µm absorption band, the broad absorption feature of CH$_4$ around the 2.3 µm region is less intense (also the shoulder of 2.0 -µm is much lower compared to the 1.9-µm shoulder). We also did not see the CH$_4$ absorption band at 2.2 µm though the band is intrinsically fairly weak and might not be expected to show up when the continuum is low. Thus, we consider the absence of the 2.2 µm band characteristic to be considered a supporting argument for H$_2$O ice abundance. Note that at temperatures of 40K – relevant to Pluto – the absorption band of H$_2$O ice (with an absence of CH$_4$) at the broader 1.5-µm is stronger than the narrower 1.65- µm (Grundy & Schmitt, 1998). However, the 1.5-µm band is not prominent in the C3 unit, likely because the C3 unit belongs to some "overall similar" pixels that might not have a strong 1.5- µm band, and also since our spectra are an average of all pixels, the 1.5- µm band can likely be subdued. Though the presence of a 1.65-µm band can be indicative of crystalline H$_2$O, an absorption band at the same wavelength is also attributed to CH$_4$ ice (Cruikshank et al., 2000). Thus, the interpretation of H$_2$O in this study considers the inspection of many aspects (e.g., spectral shape, shoulders, etc.) of absorption bands at different wavelengths (1.65, 1.9, 2.0, 2.2, 2.3 µm). Accordingly, the C3 unit is designated to have an abundance of H$_2$O ice (mixed with CH$_4$). The spatial distribution of #3 indicates that the H$_2$O ice-dominated areas are distributed at different geographic locations and associated with a variety of geologic features. Here, we are interested to probe if there are any differences in H$_2$O ice distribution due to geographic locations or if differences are more aligned to geology. To this end, we apply a post-classification scheme to distinctly separate possible H$_2$O ice-abundant subunits.

We employ the same GMM algorithm for the post-classification of the surface units (# 3 and 6) into two subunits. The probabilities of subunits for #6 are shown in the upper row and the subunits for #3 are shown in the bottom row of Fig. 6. For reference, the mean ± 1σ standard deviation of the I/F spectra from all the pixels that fall into each subunit of #3 and #6 is also included in the Appendix (Fig. A6). The post-classification result of #6 indicates that the central part of Sputnik Planitia consists of a distinct subunit while the northern part of the SP and Venera Terra share common characteristics. The Venera Terra is in the northern mid-latitudes above 50º N while Sputnik Planitia barely exceeds 45º N. The geology of the central part of SP is different from its northern part and they show different albedos (White et al. 2017). Thus, the probability plots of subunits for #6 imply that differences between $N_2$ and CO-rich areas may be controlled by latitude, geology, and/or volatile transport mechanism within the SP basin (further interpretation are given in the later section). The histogram of the probabilities extracted from the pixels of each corresponding post-class of C6 is provided in the Appendix (see Fig. A7). The histogram plot indicates that each sub-unit is dominated by pixels with higher probability values (i.e., mostly pure pixels).

Likewise, the post-classification of #3 shows that the composition ($H_2O$ mixed with other ices) at Kiladze crater is different from that of Krun Macula. The mean spectral plot (see Fig. A6 in Appendix) also indicates that there are differences in the spectral signature in those two regions most likely because the fractions of water on these units are different. The mountains and depressions (in small areal extend) in the Cthulhu Macula, Oort crater, and patchy spots in Tombaugh Regio belong to the same subunit of the Kiladze crater. However, the geographic locations of this subunit are distributed at different latitudes on Pluto's surface. Thus, the probability plot of subunits for #6 indicates that the compositional difference in $H_2O$-rich areas does not closely correlate with latitude, but rather is perhaps aligned to surface geology or other factors. We provide a histogram of the probabilities extracted from the pixels of each corresponding post class of C3 in the Appendix (see Fig. A7). The histogram plots show that the post-classified surface units are dominated by pixels with higher probability values (close to 1). This indicates dominance of mostly pure pixels at each post-class for the C3 unit.

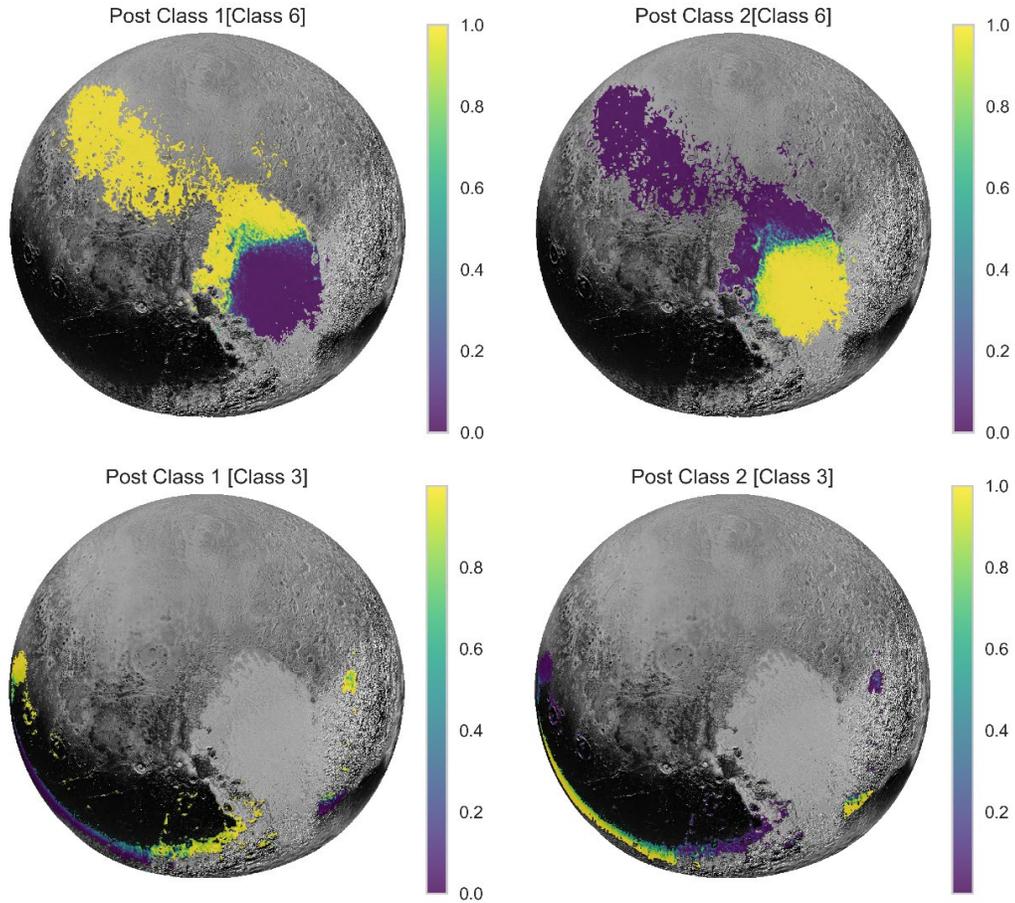

**Fig. 6:** Post-classification probabilities of subunits for the region enriched with $N_2$ and CO region (#6; upper row) and the subunits for the $H_2O$-rich areas (#3; bottom row). The central part of Sputnik Planitia shows a distinct subunit while the southern SP and Venera Terra share common characteristics. $H_2O$ ice at the Kiladze crater shows a difference from that of at Krun Macula. Refer to the histogram plot in Fig. A7 in the Appendix which shows that the post-class surface units of C3 and C6 are dominated by pixels with higher probability values.

Finally, we integrate all post-classification subunits into the original surface units shown in Fig. 4 to make a composite surface unit map encompassing a total of ten surface units (Fig. 7). In the composite surface unit map, the subunits forming the central part of Sputnik Planitia were reassigned the surface unit #9 while the subunit of Krun Macula was reassigned the surface unit #10. The compositional information of each surface unit is also labeled on the final surface unit map. Thus, we hereafter alternatively call the composite surface unit map as the generalized global compositional map. A summary of the characteristics and bulk compositions (qualitative) of the surface units is given in Table 2.

To facilitate the compositional comparison between the surface units, the average I/F spectra for ten composite surface units from the generalized global compositional map are given in Fig. 8. The colors in the average I/F spectra plot are the same as the corresponding surface unit colors in the generalized global compositional map. The characteristic absorption bands for $N_2$ at 2.15 μm and CO at 1.58 μm are labeled as the vertical lines in the plot. We also label the 1.65- and 2.0-μm absorption bands of $H_2O$. As noted above, the broad absorption bands of $H_2O$ ice are indiscernible due to overlaid strong $CH_4$ absorption bands at NIR wavelengths. Thus, the relative comparison between the shoulders at 1.9 and 2.0 μm is also considered for interpreting $H_2O$ abundance. We also utilize the relative depth (compared to other spectra) of the weak absorption band of $CH_4$ at 2.2 μm while interpreting the water-ice-dominated areas (details are mentioned above).

On the outer solar system bodies, the surface volatiles are often found in different combinations or mixtures rather than in a pure state. The abundant volatile $CH_4$ and $N_2$ ices, as well as CO ices, at Pluto's average surface temperature of 40 K, are partly soluble and can form two binary mixing phases – $CH_4$-rich ices diluted with $N_2$ ($CH_4:N_2$) and $N_2$-rich ice diluted with $CH_4$ ($N_2:CH_4$) (e.g., Prokhvatilov and Yantsevich, 1983; Trafton, 2015). Adding CO to the mix can create additional mixing phases; the phase diagram for the ternary mix is currently in development. $CH_4$ absorption bands shifts towards shorter wavelengths when a small amount of $CH_4$ is dissolved within $N_2$ such that with the increase of $CH_4$ concentration the absorption band shifts decrease (e.g., Protopapa et al. 2015). That means that higher concentrations of $CH_4$ correspond to smaller observed band shift (e.g., Scipioni et al., 2021). Thus, we use the shift of the peak at 1.69 μm as an indicator of $N_2:CH_4$ such that a shift toward shorter wavelengths represents the presence of $CH_4$ in the $N_2$ ice matrix while a shift toward longer wavelengths may be an indicator of $CH_4$-rich (with larger grains) materials.

Some of the surface units also show the dilution of $N_2$ in $CH_4$ as evident in the average spectra of the surface classes. We assess the dilution of $N_2$ in $CH_4$ ice ($CH_4:N_2$) by examining the absorption band at 1.69 μm. On Pluto, the absorption band at 1.69 μm has been historically attributed to the presence of pure $CH_4$ because the band only occurs in pure methane spectra and has not been observed in an ice mixture of $CH_4$ diluted with a smaller concentration of either α or β phases of $N_2$ ice at any temperature (Cruikshank et al., 2021). However, the recent laboratory study by Protopapa et al. (2015) demonstrated that the 1.69-µm absorption feature can no longer exclusively be attributed to the presence of pure $CH_4$ ice on Pluto's surface since the absorption feature has also been observed in $CH_4:N_2$ samples.

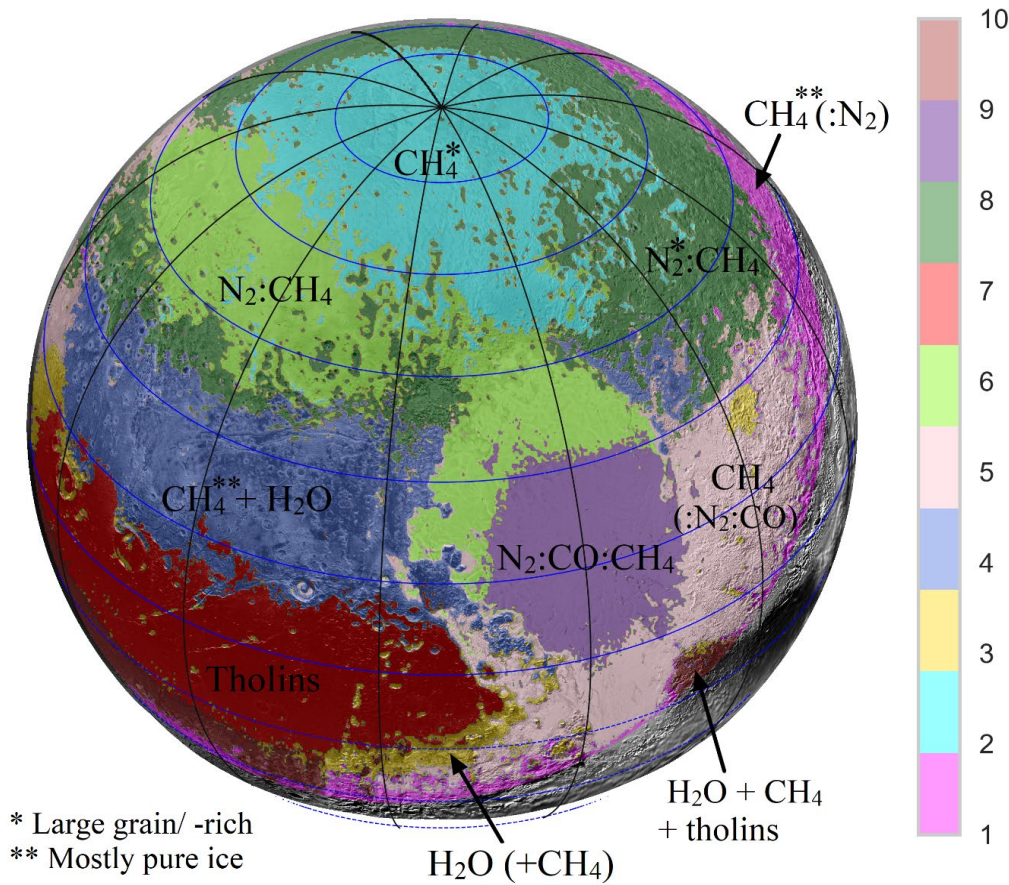

**Fig. 7:** The generalized global compositional map of Pluto superposed on a higher resolution LORRI basemap. The compositions of the surface units are labeled to the corresponding unit. The asterisk (*) superscript appended to a component refers only to that component where * = larger grain/-rich material and ** = pure ice. The latitude and longitude are labeled as the blue and black grid, respectively. The values in the legend indicate the assigned surface unit number.

**Table 2**: Summary of the characteristics and composition of the surface units.

| # Unit | Major geographic province | Composition |
|---|---|---|
| 1 | Tartarus Dorsa | $CH_4$ is diluted in a smaller amount of $N_2$; $CH_4(: N_2)$; $CH_4$ ice is mostly pure |
| 2 | Lowell Regio up to Voyager Terra to the south | Pure or larger grain of $CH_4$; perhaps no or very little $N_2$ |
| 3 | Kiladze crater, Oort crater, and Pigafetta Montes | An abundance of $H_2O$ + $CH_4$ ice; $CH_4$ snow-capped mountains supported by $H_2O$ |
| 4 | Al-Idrisi Montes, Baret Montes, Hillary Montes, Viking Terra | $CH_4$ and $H_2O$; $CH_4$ ice is mostly pure; perhaps a very little amount of $N_2$ |
| 5 | South of Sputnik Planitia and Tombaugh Regio | $CH_4$ ice might be slightly diluted with $N_2$ and CO; $CH_4(: N_2:CO)$ |
| 6 | Northern lobe of Sputnik Planitia and Venera Terra | $N_2$ ice diluted with $CH_4$ ($N_2:CH_4$); perhaps a very little amount of CO |
| 7 | Cthulhu Macula | Red materials, tholins, refractory organics [inferred from colors and not spectral bands] |
| 8 | Pioneer Terra and the base of Burney crater | $N_2$ ice diluted with $CH_4$ ($N_2:CH_4$); coarser $N_2$ ice grains |
| 9 | Central Sputnik Planitia; the heart of SP | $N_2:CO:CH_4$; the highest abundance of $N_2$ and CO; the cold trap |
| 10 | Krun Macula and a belt follow south of Cthulhu Macula | A mixture of $H_2O$ + tholins + $CH_4$; absorption bands are not stronger |

Using the global compositional maps, we focus on the broad generalized regions on Pluto's surface that show particularly interesting comparisons between their spectral composition with respect to geologic structures and topographic maps. Previous studies using MVIC (Multispectral Visible Imaging Camera) also noticed a latitudinal pattern to some of Pluto's ices (see Earle et al., 2018). We highlight these interesting features on Pluto where different volatile interactions are occurring and demonstrate the level of detail between spectral and geological study at Pluto. Note the surface unit we refer to hereafter in our discussion corresponds to the surface units in generalized global compositional maps in Fig. 7 and corresponding average I/F spectra in Fig. 8. A list of the spectral components used to characterize surface compositions and surface units that exhibit the spectral components is also provided in table 3. We also incorporate an interpretation of the geology of the surface units based on existing literature.

*Area 1: Sputnik Planitia (SP)*

SP is an $N_2$-dominated, volatile-filled basin that is currently undergoing solid-state convection, creating large (10 – 40 km across) polygonal cells across the basin (Stern et al., 2015; Protopapa et al., 2017). SP has also been observed from MVIC data to have minor concentrations of methane, with a weaker **strength in** MVIC equivalent width of **890 nm absorption band** (Earle et al., 2018). The composite surface unit map shows that Sputnik Planitia exhibits multiple classes with different surface compositions. Notably, the central part has spectral signatures of a higher concentration of $N_2$ and the presence of CO ices, while the northern portion has a lower concentration of $N_2$. Overall, the SP basin appears to have three main spectral regions – corresponding to I/F spectra indicated as C5, C6, and C9 (see Fig. 8).

The northwestern-most lobe of SP is darker in albedo (White et al. 2017). This portion of SP is indicated as spectrum C6, which has a small $N_2$ signature and no discernible (or very weak) CO feature. The peak of the 1.69-µm band slightly shifts towards shorter wavelengths – representing a surface composition of $N_2$ ice diluted with $CH_4$ ($N_2$:$CH_4$). As previously discussed, when $CH_4$ is diluted in $N_2$, there is a shift of the $CH_4$ bands to slightly shorter wavelengths (Quirico and Schmitt, 1997). The unit may host a very low concentration of CO since there is no discernible (or very weak) absorption at 1.58 µm. From the geologic mapping of SP (White et al., 2017), this part of SP has dark-cellular plains (*dcp*) and dark-trough-bounding plains (*tbp*). The *dcp* unit is deficient in $N_2$ relative to the innermost (central) portion of SP, also displaying a higher content of entrained tholins, along with the *tbp* unit (Scipioni et al., 2021). The *tbp* unit also represents entrained tholins at the edges of the convective cells and bounding troughs.

The central lobe of SP has the spectral indicator C9, which is observed to have a more prominent signature of CO and $N_2$ compared to the other spectra. Having the presence of $N_2$ and CO in this central portion of SP gives more evidence into SP being a cold trap of such volatiles (Bertrand & Forget 2016; Keane et al. 2016). Moreover, the peak at 1.69 µm shows its highest shift towards shorter wavelengths compared to other spectra – representing the presence of the highest concentration of $N_2$ ice at the central part of SP compared to all other units. The CO absorption at 1.58 µm also shows a sharp feature. This represents a surface unit of the highest amount of $N_2$ and CO deposition. The geologic unit corresponding to this part of SP is bright-cellular-plains (*bcp*). The *bcp* unit is inferred to be resurfaced (relatively fresh, bright) $N_2$ ice (also inferred from the lack of impact craters in this region; see Singer et al. 2021). The cellular patterns indicate active solid-state convection (McKinnon et al. 2016). The higher albedo could be the condensation of $N_2$ from the atmosphere (White et al., 2017).

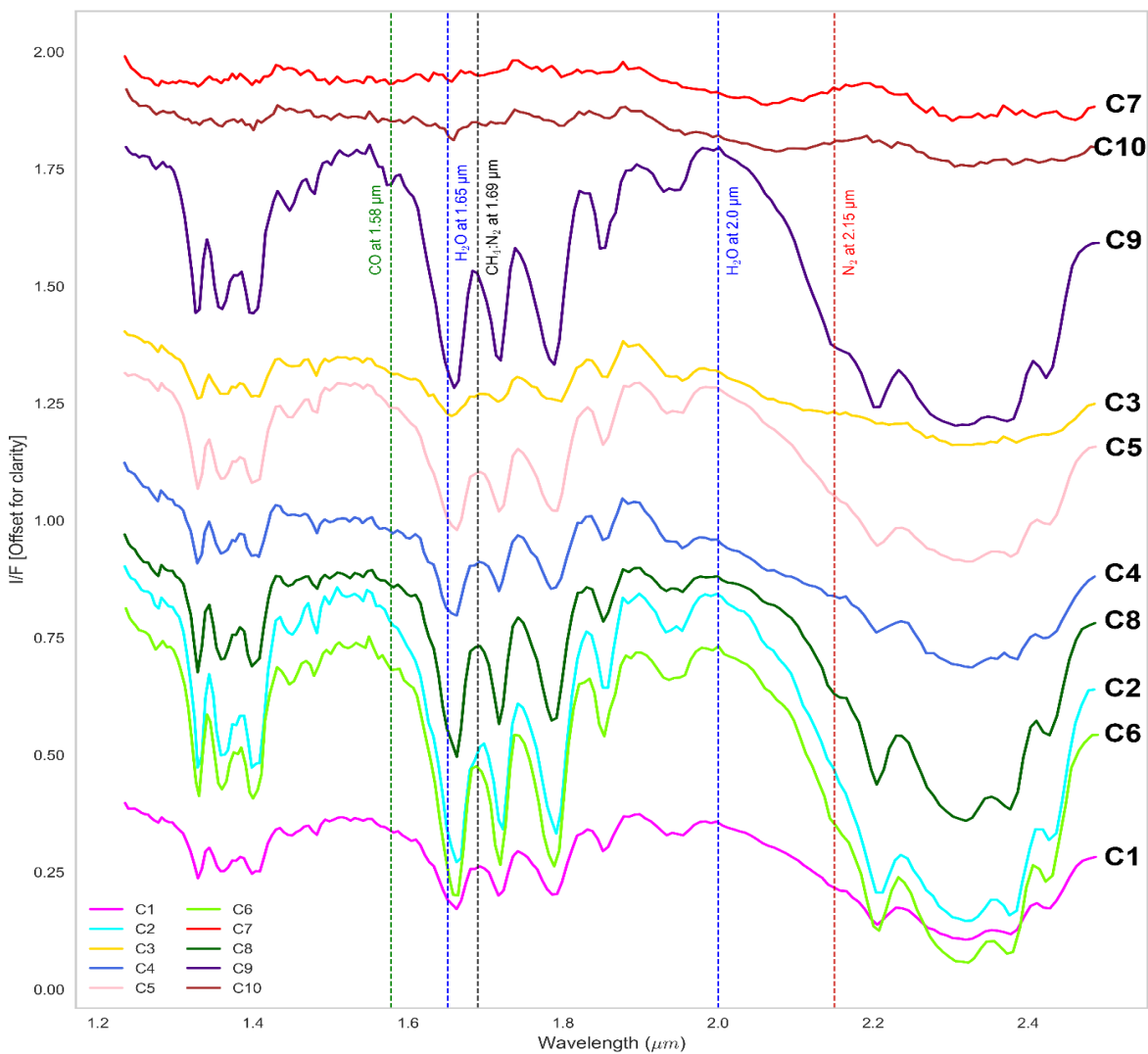

**Fig. 8:** The average I/F spectra of the ten composite surface units. The colors of the spectra correspond to the colors of the surface unit in Fig. 7. The dashed vertical lines indicate $N_2$ absorption at 2.15 μm (red), CO absorption at 1.58 μm (green), $CH_4:N_2$ absorption at 1.69 μm (black), and $H_2O$ absorption at 1.65 and 2.0 μm (blue). Note that the I/F values in the y-axis have offsets for clarity. For a higher resolution colored version of the figure, the readers are referred to the online version of the paper[3].

---

[3] The wavelengths and average I/F data in the figure are available in the publicly available repository at https://github.com/alemran042/Pluto-PC-GMM-Data

The southern-most unit within the SP basin (along with Tombaugh Regio) is spectral unit C5, which is observed to have a smaller spectral signature of $N_2$ and CO. The $N_2$ observed in this surface unit could be from the lightly pitted plains (*lpp*), in which $N_2$ is sublimating from the pits (White et al., 2017). This region also has patchy, pitted, marginal plains (*pmp*), which have shallow $N_2$ ice, and sparsely pitted plains (*spp*), which host $N_2$ ice, either from resurfacing or atmospheric condensation (White et al., 2017).

**Table 3**: List of the spectral components used to characterize compositions and their surface units.

| Composition | Spectral component (µm) | Surface units (#) |
|---|---|---|
| $N_2$ | 2.15 µm | C8, C9 (strong) <br> C1, C5, C6 (weak) |
| $CH_4$ | 1.30 – 1.43 µm, 1.59 – 1.83 µm, 1.90 – 2.0 µm, and 2.09 – 2.48 µm | All units with different strengths, except C7 |
| CO | 1.58 µm <br> 2.35 µm (indiscernible) | C9 (strong) <br> C5, C6 (weak) |
| $H_2O$* | 1.65 µm, <br> 2.0 µm (shoulder) | C3, C4, C10 |
| $N_2$:$CH_4$ | peak shift at 1.69 µm <br> (blue or red shift) | C6, C9 (blue shift) <br> C2 (redshift) |
| $CH_4$:$N_2$ | absorption at 1.69 µm | C1, C4 (strong) <br> C3 (weak/ indiscernible) |

*Note:* We also consider the shape of the characteristic absorption band of $CH_4$ at 2.2 µm for the interpretation of the presence of $H_2O$ ice. Please see the text for detailed description of the approach.

*Area 2: Al-Idrisi Montes and western SP glaciers*

The Al-Idrisi Montes and the western shores of SP show an amalgamation of chaotic blocky terrain and mountainous glaciers, such as the Baret Montes and Hillary Montes glacier chain to the southwestern shores, rising several kilometers over SP (Moore et al., 2016). These mountains and glaciers are reddish in nature, while the chaos blocks to the northwestern portion of SP are more neutral in coloring. The glaciers and chaotic terrain floating on the western shores of SP are mainly water ice blocks (Howard et al. 2017; Cook et al. 2019; White et al. 2021). This spectral region (indicated as C4) from our global maps also extends in a latitudinal band into Viking Terra [~13° - ~40°N], tracing the northern boundary of Cthulhu Macula.

From the spectra, the $CH_4$ signature is not as deep comparatively, probably indicating some dilution in $N_2$ ice (Schmitt et al., 2017). However, interestingly, a slight 1.69-μm absorption band is seen in the spectra (C4). However, we do not see any discernible $N_2$ absorption band at 2.15 μm. As has been noted above, the 1.69-μm absorption band is attributed to the presence of both pure $CH_4$ and $CH_4:N_2$ ice (Protopapa et al., 2015; Cruikshank et al., 2021). Thus, based on spectral characteristics we presume the composition of the surface unit #4 to be the presence of pure $CH_4$ ice. The $H_2O$ correlation map by Cook et al. (2019) also reported the presence of water ice throughout this surface unit. Though the average spectra (C4) do not clearly show a strong absorption band at 1.65 μm, the shoulder peak at 1.9 μm is much higher compared to 2.0 μm. Thus, the possible composition for this surface unit is $CH_4$ + $H_2O$. We note that this latitudinal band does not extend to the eastern close-encounter hemisphere (except for a very small area on the northeastern shores of Sputnik Planitia).

*Area 3: Northern Terrae and Polar Cap*

The northern plains of Pluto are sectioned into different major regions plus the polar cap. The polar unit (officially known as Lowell Regio) is predominantly our spectral unit C2, which is a sampling of $CH_4$-rich components with little or no detection of CO and $N_2$. Notably, the peak of 1.69 μm shows a shift toward longer wavelengths among other spectra, indicating the presence of a $CH_4$-rich component. This pure $CH_4$ ice was also confirmed by observations from Schmitt et al. (2017). Venera Terra [Lat 38° – 74.9°N; Lon 99.8° – 135°E] is dominantly the spectral unit C6, which consists of $CH_4$ absorption bands and with a weak spectral signature of $N_2$. Like the northern portion of SP, this region has a surface composition of $N_2:CH_4$ ($N_2$ ice diluted with $CH_4$). Venera Terra geology involves the large (585-km long) Djanggawul Fossae and Piri Rupes structure (549-km long).

Pioneer Terra [Lat 43.5° - 69°N; 170° – 214°E] is dominated by unit C8, which is mostly $CH_4$ and $N_2$ mixtures ($N_2:CH_4$). However, compared to the Venera Terra unit (C6), this unit has a much broader 2.15-μm absorption band which is probably indicative of the presence of coarse grain $N_2$ ice. The Hayabusa Terra [Lat 28° – 62°N; 196° – 264°E], the easternmost region in the close-encounter hemisphere, is also dominated by the C8 spectral unit. Within these icy compounds, $N_2$-rich ices can be very coarse-grained (few to 10s of centimeters), whereas $CH_4$-rich ices are typically < 1 mm in size (Protopapa et al. 2017). Most of the region is speckled with large, flat-floored pits and uplands (Howard et al. 2017). Water ice forms the base layer bedrock but is extensively mantled by other icy constituents. These erosional regions may reflect bulk composition, though the mantles may be of deeper, varied depths that may reflect seasonal volatile interactions and migrations (Earle et al. 2017; Protopapa et al. 2017). While $CH_4$ tends to accumulate at higher elevations, $N_2$ collects in depressions (Bertrand and Forget 2016). $N_2$ ice detected in the spectral unit C8 is compatible with the origin of $N_2$ by atmospheric condensation condensing in these low-lying depressions.

The northwestern-most close-encounter hemisphere region is Vega Terra [Lat 10.5°– 57°N; Lon 42.9° – 128°E]. There are two main spectral units from our global maps observed in this region: C6 and C8. C6 is in the northern portion of Vega Terra, which is observed to have a small amount of $N_2$, like the northern portion of SP. The southern part of Vega Terra is a mixture of $CH_4$ and $N_2$ (C8) in a much narrower latitudinal band. The geology in this portion of Pluto is mostly $CH_4$-haloed impact craters (Villaça et al. 2021).

Voyager Terra [Lat 36° – 77°N; Lon 129° – 176.5°E] consists of blotches of the spectral units C6, C2, and C8. We note that the C8 unit also corresponds to the relatively higher elevations of Hunahpu Valles, which has also been verified with observations of $CH_4$ concentrations at higher elevations in the MVIC data (Earle et al. 2017). These concentrations of $CH_4$ and $N_2$ could correspond to the condensation of atmospheric volatiles, like the $CH_4$-capped mountains observed in Cthulhu Macula (Bertrand et al. 2020).

### *Area 4: Cthulhu Macula and Southern Tombaugh Regio*
Cthulhu Macula is a large, dark, reddish region in Pluto's western close-encounter hemisphere. The reddish coloring is possibly mostly the result of atmospheric tholins deposition (see Grundy et al. 2018). However, the reddish-colored component at Virgil Fossae has been hypothesized to originate from the recent cryovolcanic activities at the fossae (Cruikshank et al., 2019). This region shows little to no methane absorption, as also confirmed by previous MVIC studies (Earle et al. 2018). To the southeast of Cthulhu Macula (southern portion of Tombaugh Regio) lies the putative cryovolcanic structures, Wright Mons and Piccard Mons (Singer et al., 2022). While most of the region consists of spectral units C3 and C5, it is interesting to note that the spectral unit C1 intersects the summit pit depression of Wright Mons, and continues at an angle through Krun Macula and Tartarus Dorsa in the close-encounter eastern hemisphere.

The C1 unit is very similar to the C5 spectra, though the C1 may have a small detectable amount of $N_2$ that the C5 unit doesn't exhibit. Interestingly, C1 shows the presence of a 1.69-μm absorption band coupled with the $N_2$ absorption at 2.15 μm. The 1.69-μm absorption has indeed been attributed to pure $CH_4$ or $CH_4$:$N_2$ (Protopapa et al., 2015) which the C1 spectrum corroborates. Thus, the likely composition at C1 is that $CH_4$ ice might be slightly diluted with $N_2$ [$CH_4$(: $N_2$)]. The detection of $CH_4$ at Wright Mons most likely represents a thin surficial layer deposited out of the atmosphere (Singer et al., 2022). $N_2$ and $CH_4$ have been previously observed at Tartarus Dorsa as the main drivers of the sublimation processes of the penitents (Moores et al., 2017).

*Area 5: Water ice areas*

The spectra for the Kiladze crater unit (C3) show evidence of abundant water ice. There is a clear presence of the $H_2O$ absorption band at 1.65 μm. Though the 2.0-μm absorption band is not strongly evident in the spectra, the shoulder at 2.0 μm is much lower than 1.9 μm. We label this surface unit as the crystalline water ice abundant area, consistent with the previous study by Cook et al. (2019). We do not see the presence of a characteristic $CH_4$ absorption band at 2.2 μm, though other absorption bands of $CH_4$ are weakly present. Thus, the $H_2O$ ice in the Kiladze crater area likely coexists with $CH_4$ ice ($H_2O$ + $CH_4$ ice). The unit is also evident in the mountains and small depression in Cthulhu Macula. In the case of Krun Macula, neither $CH_4$ nor $H_2O$ has strong absorption bands. The spectra show a similar shape to the spectra of Cthuhulu Macula – an indication of the presence of red materials or tholins. We note that the presence of tholins and other red materials is inferred on the basis of spectral slope (color), both in the spectral region covered by LEISA and analyzed here, and by MVIC at shorter wavelengths, and not on discrete, assignable spectral bands. Thus, on the basis of spectral shape, we infer that the surface unit at Krun Macula is likely to host a mixture of $H_2O$, $CH_4$, and tholins.

**Discussion**

The PC-GMM method implemented in this study not only maps the geographic distribution of surface units on Pluto but also distinguishes the differences in composition (both mixing and physical states of volatiles) from the average I/F spectra of the corresponding surface units. The generalized compositional map shows the presence of different surface units at different geologic features. The mountains such as Pigafetta and Elcano Montes in the Cthulhu Macula region (Fig. 9a) belong to a surface unit comparable to the Kiladze crater area and their spectra show the presence of an $H_2O$ absorption band mixed with $CH_4$ ice. The top of these mountains hosts a high concentration of methane while the lower altitudes are mostly $CH_4$ depleted (Earle et al., 2018; Bertrand et al., 2020). The undulating mountainous terrains on Pluto suggest that $H_2O$ ice forms the supporting bedrock for the surface volatiles and holds the base of the mountains (Stern et al., 2015; Spencer et al., 2020). Thus, our result bolsters the hypothesis that they are $CH_4$-capped $H_2O$-supported mountains. The presence of an $H_2O$-rich surface unit on one side of the Kiladze crater (Fig. 9b) may be indicative of the spatial span of the crater ejecta.

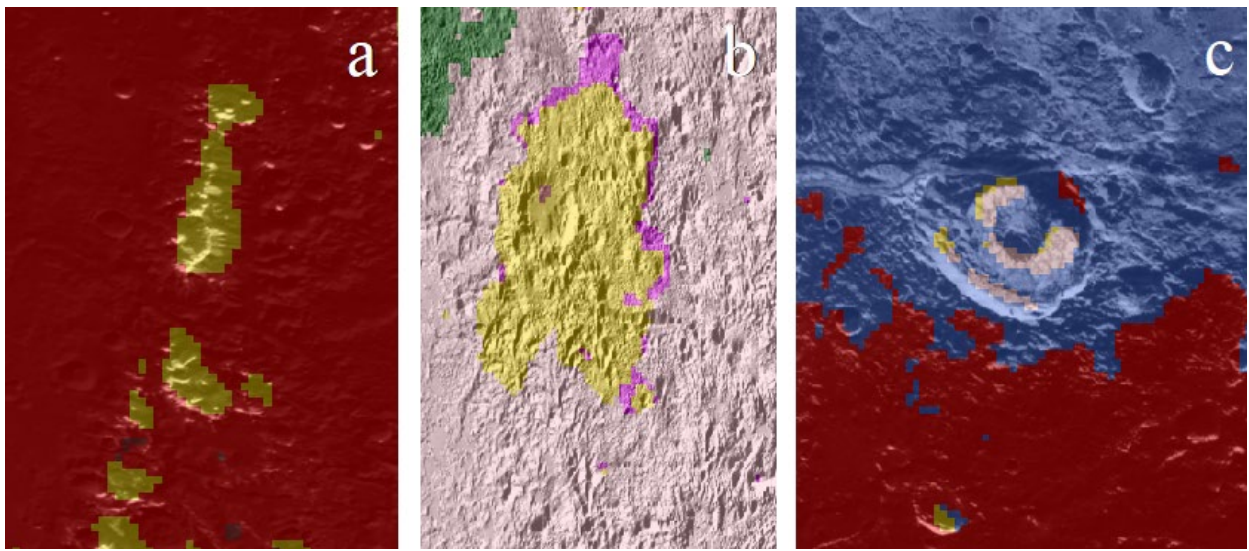

**Fig. 9:** The subset of the generalized surface map at mountains in the Cthulhu Macula (a), Kiladze crater (b), and Elliot crater (c). For the reference location, please see Fig. 1a.

The PC-GMM method also performed well in distinguishing the local scale difference in surface composition. The base of the Elliot crater (Fig. 9c) shows the presence of multiple surface units. The bottom floor indicates that surface unit #5 follows the central peak of the crater (except the north-eastern part of the crater floor). This unit corresponds to a surface composition rich in $CH_4$ ice which might be slightly diluted with $N_2$ and CO. It is notable that this surface unit (#5) shows no $H_2O$ ice. The local scale compositional study at Elliot crater also confirmed the presence of non-$H_2O$ materials at these locations (Dalle Ore et al. 2019). A few pixels on the floor of the crater show # 3 – presence of crystalline $H_2O$ ice mixed with $CH_4$. The presence of the $H_2O$ ice at these isolated spots on the floor of the crater is also consistent with Dalle Ore et al. (2019).

The **areas with $H_2O$ ice** (i.e., #3, #4, and #10) show an interesting distribution across Pluto since these surface units are limited below ~45ºN. We did not find any water ice-containing surface unit from ~45ºN to the north pole. The absence of exposed $H_2O$ in the higher northern latitude can be explained by the thick coverage of the volatile deposits at these latitudes (Cruikshank et al., 2019). This hypothesis is consistent with the results of our study such that of the distribution of the $CH_4$-rich unit (#2) on the north pole surrounded by the $N_2$:$CH_4$ units (#6 and #8) on the Venera Terra, Voyager Terra, and Pioneer Terra. We found $H_2O$-rich areas in some small craters (depressions) at Cthulhu Macula or associated with specific geologic features. For instance, the presence of the $H_2O$ ice-rich surface unit (#3) was seen along the Virgil Fossae, Beatrice Fossae, Edgeworth crater, and Oort crater areas. This suggests that the exposed $H_2O$ distribution on Pluto's surface is likely controlled by geologic features.

The map of the global compositional distribution of surface ices is largely consistent with the compositional mapping by previous studies by Schmitt et al. (2017) and Protopapa et al. (2017). On top of that, the current study provides a quantitative aspect of the surface units thanks to the ability to estimate the areas of the different segments in the Sputnik Planitia basin. The central part of SP – considered as the cold trap for $N_2$ and CO ices – has an estimated area of ~284,000 sq. km which is more than one-third (1/3) of the total area of the basin. Based on our global-scale compositional distribution result we can infer volatile transport directions across the dwarf planet (Fig. 10) confirming and refining the results from the study of Protopapa et al. (2017) as mentioned in Cruikshank et al. (2021).

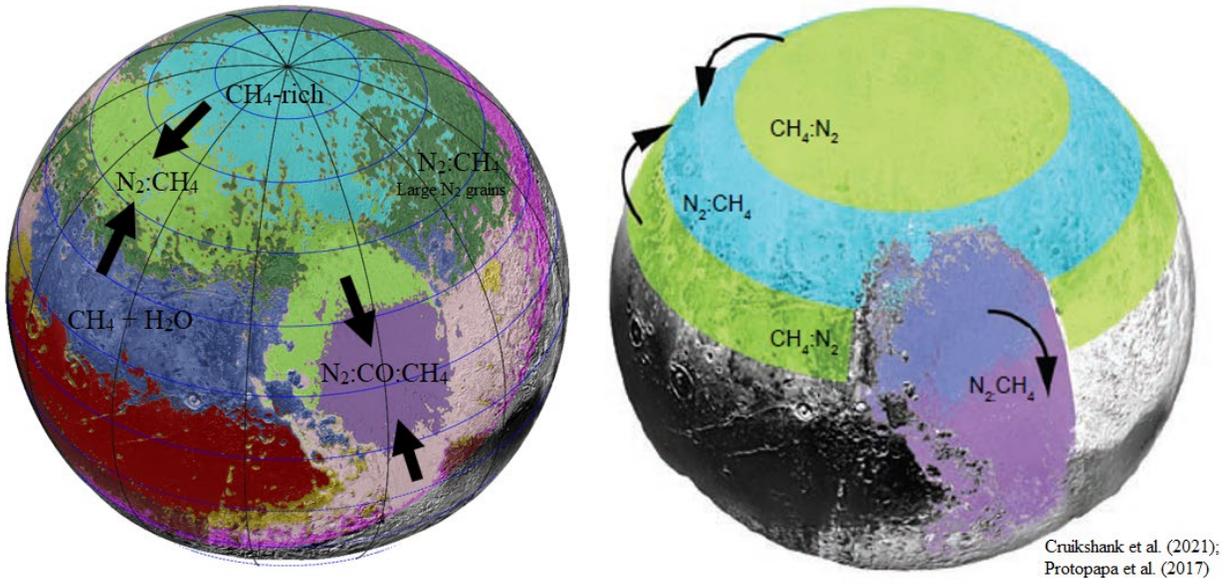

**Fig. 10:** Global-scale surface compositional distribution and inferred volatile transport directions (indicated by arrows) proposed in this study (a) and schematic view of the major ice distribution and direction of nitrogen sublimation transport (b) explained by Protopapa et al. (2017). Figure (b) was adopted from Cruikshank et al. (2021) and Protopapa et al. (2017).

Non-volatile $H_2O$ ice should be exposed on the surface unless it's covered by the seasonal deposits and substantial reservoirs of the $CH_4$ and $N_2$ ices (Schaller and Brown, 2007). Exposed $H_2O$ ice has not been detected across all the latitudes except in small patchy areas. The latitudinal patterns in the volatile $N_2$, $CH_4$, and CO distribution across Pluto's surface might suggest that insolation is the major controlling parameter for volatile distribution (Protopapa et al., 2017). However, the distribution of surface compositional units also corresponds to larger geological provinces (e.g., Venera Terra and Pioneer Terra) and local scale topography (e.g., Pigafetta Montes, Baret Montes, Elliot crater). Thus, we presume that besides being influenced by insolation, the distribution of volatile ices on Pluto's surface is controlled by geological structure and topography – consistent with Protopapa et al. (2017) and Bertrand et al. 2016.

Among the $N_2$, $CH_4$, and CO ices, $N_2$ is the most volatile component; it has the highest vapor pressure and sublimates first when heated by solar irradiation. Thus, the spatial abundance and distribution of nitrogen on Pluto's surface highlight the direction of nitrogen sublimation transport or volatile transport. A volatile transport (nitrogen sublimation and deposition) mechanism is interpreted for the northern hemisphere since the compositional surface units follow latitudinal patterns. The vigorous spring sublimation after a long polar winter over the past 20 years may lead to the illuminated north pole being devoid of $N_2$ ice by sublimation and redeposition at latitudes southward (Protopapa et al., 2017). However, between the years 1975 to 1995, the region covering surface unit #4 (i.e., northern lower latitudes) received more solar heating than the north (Protopapa et al., 2017). The intense solar heating at this latitude resulted in the sublimation of $N_2$ from these areas, which was then atmospherically transported and eventually deposited in the north (Protopapa et al., 2017). Consequently, sublimated $N_2$ ice from the north pole and northern lower latitudes found its resort in the regions between these surface units. Notably, the strength of the $N_2$ absorption band at 2.15 μm increases from the north pole (#2) southward (#8). This indicates the decreasing relative abundance of $CH_4$-to-$N_2$-rich ices with decreasing latitudes from the north pole southward (Cruikshank et al., 2021).

The **Sputnik Planitia** basin exhibits a distinct pattern of volatile transport by itself which, in fact interrupts the latitudinal patterns of volatile distribution and $N_2$ sublimation transport described above. The central part of the Sputnik Planitia hosts a higher abundance of the $N_2$ (and CO) ice compared to the northern and southern lobes of the basin. Thousands of elongated dark-floored pits structures are seen in higher resolution LORRI images at the southern lobe of the SP. Evidence of mass loss and sublimation erosion of the dark-floored pits structures in the southern lobe of SP has been observed (Stern et al., 2021). This mass loss by these pits has been hypothesized to be deposited elsewhere on Pluto's surface through a volatile transport mechanism (atmospheric transport). Since this missing mass cannot reside in the atmosphere it has to be transported and re-condensed elsewhere (Stern et al., 2021). Our results support that the sublimation mass losses of these pits are deposited on the central part of SP (indicated by the black arrow shown in Fig. 10). However, this postulation of volatile migration at SP is inferred from the simple interpretation based on geological evidence (Stern et al., 2021) and correlation between the adjacent compositional units and does not account for the effects of insolation and winds.

The **northern lobe of SP** has a shallower 2.15-μm absorption band of $N_2$ compared to the central part. This suggests that the northern lobe has a lower abundance of $N_2$:$CH_4$ while the **central part of SP** has a much smaller concentration of $CH_4$ in $N_2$ (Cruikshank et al., 2021).

Protopapa et al. (2017) interpreted these characteristic differences to be the transport of $N_2$ from the northwestern lobe southward. Loss of ices in northwestern SP (on both annual and seasonal timescale) and redeposition of the ices below 25º N have also been described by Bertrand et al. (2018). The sublimation of $N_2$ at northwestern SP leaves behind $CH_4$ as a lag deposit (Cruikshank et al., 2021). Thus, we presume that the **northern** and **southern lobes of SP** may host active $N_2$ sublimation, whose sublimated mass is then atmospherically transported and ultimately deposited in the **central part of SP**. In recapitulation, we illustrate the source and sink of the $N_2$ ice sublimation transport (volatile transport mechanism) across the Sputnik Planitia basin.

We applied the PC-GMM technique to the LEISA data on Pluto's encounter disk of the 2015 New Horizons' flyby. The results showed a latitudinal pattern in the volatile ices' distribution and showcase the nitrogen transport sublimation patterns at Pluto's global and local scale in the case of Sputnik Planitia. On the planetary scale (Pluto encounter hemisphere during New Horizons flyby), the mapping result of this study showed convergence with the volatile transport— and global circulation— model on the dwarf planet (e.g., Protopapa et al., 2017; Forget et al., 2017; Cruikshank et al., 2021). We emphasize that a similar method can be applied to other planetary bodies to characterize their surface compositions on a global scale. However, the results from our unsupervised clustering (GMM) are also consistent with the results of local scale applications of different other clustering techniques (i.e., K-means, Dalle Ore et al., 2018). Thus, we don't rule out the fact that an application of other unsupervised learning such as K-means or spectral clustering may render a satisfactory result in mapping the surface composition of planetary bodies.

**Conclusions**

Mapping spatial abundance and geographic distribution of ices and their relationship with the geological features are crucial for an in-depth understanding of volatile transport on Pluto (Bertrand and Forget, 2016). Thus, the surface compositional abundance and geographic distribution of volatile ($N_2$, $CH_4$, and CO) and non-volatiles ($H_2O$, tholins, etc.) have been thoroughly reported in the existing studies from near-infrared spectral observations from the instruments onboard New Horizons. However, existing studies in mapping surface ices on Pluto were mostly accomplished using band depth measurements of known ices, spectral indices or indicators, or implementation of radiative transfer models. The execution of these methods requires prior knowledge of representative surface ices or needs the reflectance or optical constants data of the end members (label data). Thanks to the development of sophisticated machine learning algorithms, unsupervised learning methods have gained popularity due to their agnostic wide applicability and rendering of satisfactory results.

We use the Gaussian mixture model, an unsupervised machine learning technique, to map the geographic distribution of ices on Pluto's surface. The input hyper-dimensional data were reduced to the lower dimension without compromising a substantial variation of the original data, followed by an implementation of GMM. The average I/F spectra from each surface unit have been extracted and analyzed – in terms of position and depth of the major absorption bands of $CH_4$, $N_2$, CO, and $H_2O$ – to connect the surface units with latitudes and geologic features. The geographic distribution of the surface units demonstrates a correlation with latitudinal patterns and geologic features. The method was able to recognize local-scale variations (interpreted as indicative of surface composition and the physical states of ice) of the geological features. The mapped distribution of surface units and their compositions are consistent with the existing literature and help in an improved understanding of the volatile transport mechanism on Pluto, both on global (encounter hemisphere of Pluto during New Horizons flyby in 2015) and local scales.

We implement the unsupervised method without any label data or optical constants and showed a satisfactory result. Thus, an application of this unsupervised learning technique can specifically be beneficial for mapping the distribution of surface constituents of a planetary body when label data are poorly constrained or completely unknown. An improved understanding of the surface volatile transport mechanism at planetary scale can be attained through mapping of the geographic distribution of surface material. Thus, we emphasize that the unsupervised learning (PC-GMM) used in this study has wide applicability and can be expanded to other planetary bodies of the Solar System for mapping surface material distribution and achieving a complete understating of volatile transport modeling in particular and physical processes in general on the icy bodies at the outer solar system.

**Acknowledgment**

The authors would like to thank The National Aeronautics and Space Administration (NASA)'s New Horizons mission team for collecting and processing LEISA/RALPH instrument data. We use the derived data (I/F) products of LEISA scenes generated by the New Horizons Pluto encounter surface compositional science theme team available at NASA Planetary Data System: Small Bodies Node (https://pds-smallbodies.astro.umd.edu/). The authors would like to thank Will Grundy, Leslie Young, Richard Binzel, Jason Cook, Francesca Scipioni, Silvia Protopapa, and anonymous reviewers for their useful comments.

**Appendix:**

**Section A.1.** Cumulative explained variance and scree plot of PC axes

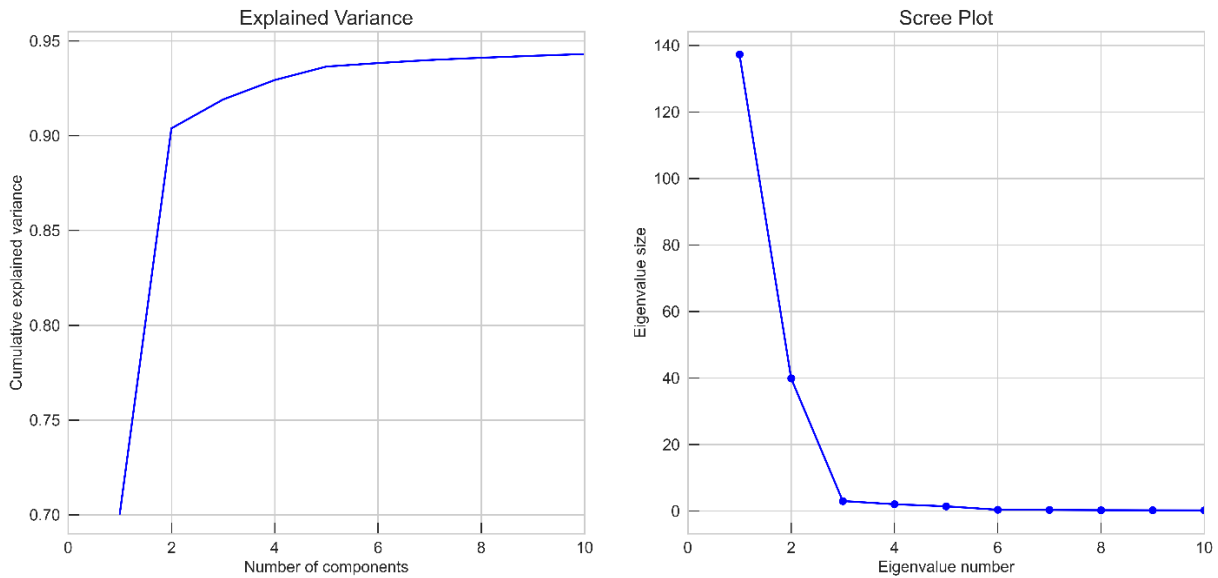

**Fig. A1:** The cumulative explained variance (left panel) and scree plot (right panel) of the PC axes. At four pc-axes, the plot shows 92.93% total variance of the data. The pc#5 and onward the plot shows a very gentle increase in the cumulative explained variance.

**Section A.2.** EM Algorithm for Multivariate Gaussian Mixture Model

---

*Algorithm:* EM Algorithm for Multivariate Gaussian Mixture Model

**Initialize:** $\pi_j^{(0)}$, $\mu_j^{(0)}$ and $\Sigma_j^{(0)}$ for $j = 1, 2, \ldots, k$.

**for** $m = 1, 2, \ldots$ **do**

**E-step:** Let $\theta^{(m)} = \{\mu_1^{(m)}, \mu_2^{(m)}, \ldots, \mu_k^{(m)}, \Sigma_1^{(m)}, \Sigma_2^{(m)}, \ldots, \Sigma_k^{(m)}, \pi_1^{(m)}, \pi_2^{(m)}, \ldots, \pi_k^{(m)}\}$ is the set of $\theta$ at $m$-th iteration. For $i = 1, 2, \ldots, n$ and $j = 1, 2, \ldots, k$ compute

$$\gamma_{ij}^{(m)} = p(z_i = j | X_i, \theta^{(m-1)}) = \frac{\pi_j^{(m-1)} N_d\left(X_i | \mu_j^{(m-1)}, \Sigma_j^{(m-1)}\right)}{\sum_{j=1}^{k} \pi_j^{(m-1)} N_d\left(X_i | \mu_j^{(m-1)}, \Sigma_j^{(m-1)}\right)} \quad (6)$$

$\gamma_{ij}^{(m)}$ is the posterior probability that $z_i = j$ given $X_i$ and $\theta^{(m-1)}$ at $m$-th iteration.

**M-step:** For $j = 1, 2, \ldots, k$ compute

$$\mu_j^{(m)} = \frac{1}{n_j^{(m)}} \sum_{i=1}^{n} \gamma_{ij}^{(m)} X_i \quad (7)$$

$$\Sigma_j^{(m)} = \frac{1}{n_j^{(m)}} \sum_{i=1}^{n} \gamma_{ij}^{(m)} (X_i - \mu_j^{(m)})^T (X_i - \mu_j^{(m)}) \quad (8)$$

$$\pi_j^{(m)} = \frac{n_j^{(m)}}{n} \quad (9)$$

where $n_j^{(m)} = \sum_{i=1}^{n} \gamma_{ij}^{(m)}$

**end for.**

---

**Section A.3.** Scaled AIC and BIC values at the different # of clusters

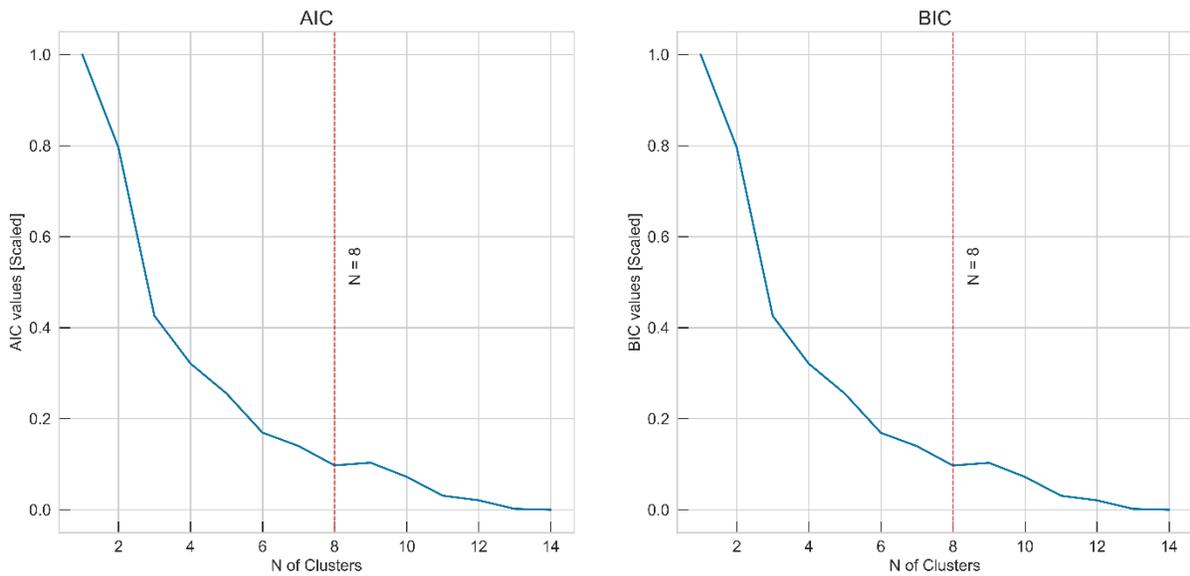

**Fig. A2**: Scaled AIC and BIC values at the different number of clusters. Both AIC and BIC show a similar trend as a function of the number of clusters. The plot shows the first local minima BIC (and AIC) value to be 8 and therefore was considered the optimal # of clusters in this study.

**Section A.4.** Linear fits between the observed and predicted density estimates.

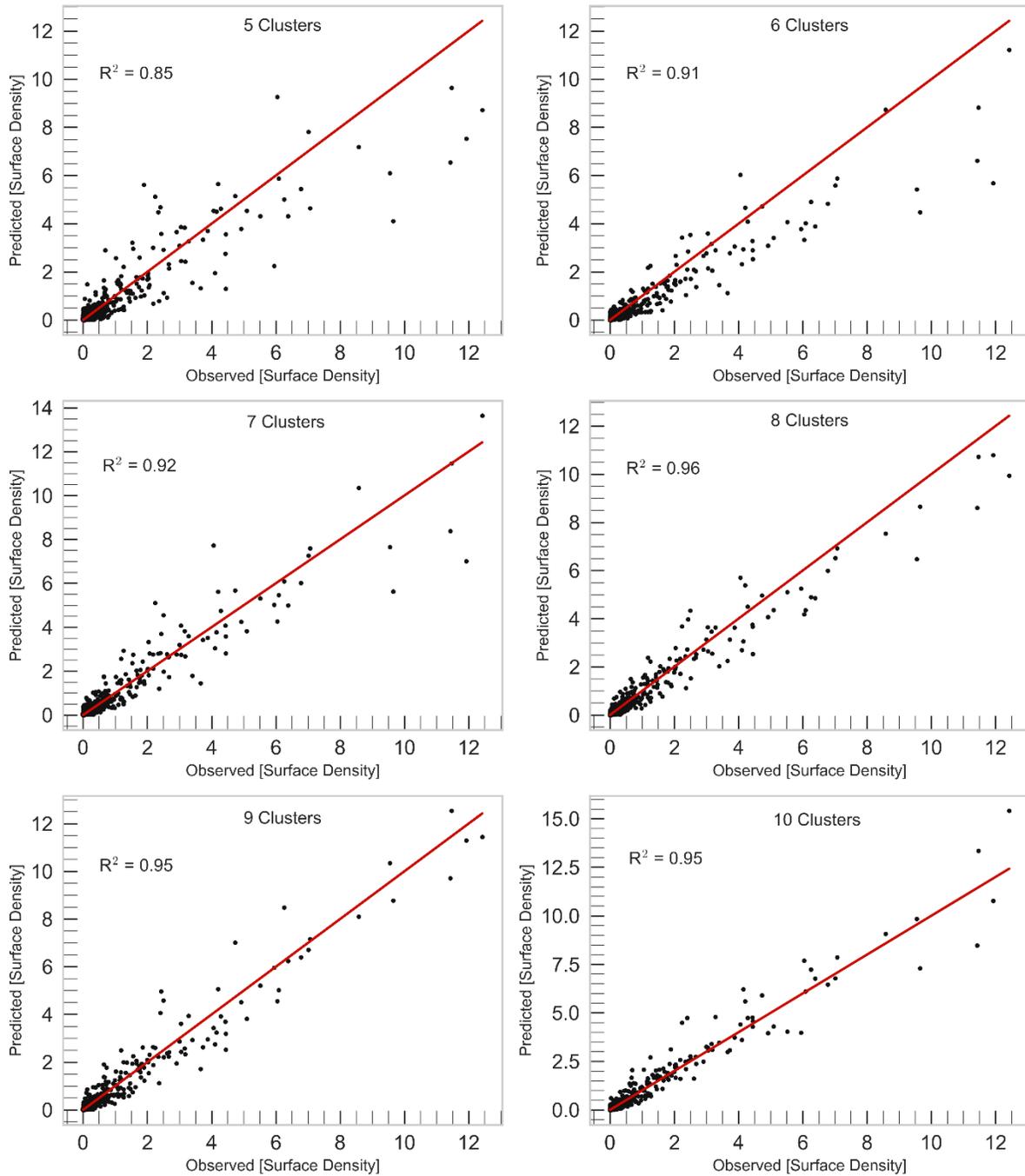

**Fig. A3:** Observed and predicted density plot for different numbers of clusters using the PC-GMM algorithm. The observed data were the first 4 pc-axes used as the input data for GMM while the predicted data were simulated from the resulting GMM model parameter. The density for both observed and predicted data was calculated from multivariate kernel density estimates. The coefficient of the determination ($R^2$) for linear fit between the observed and predicted density for 8 GC solution is 0.96 – suggesting a good fit of the model.

**Section A.5.** Pluto Nomenclature.

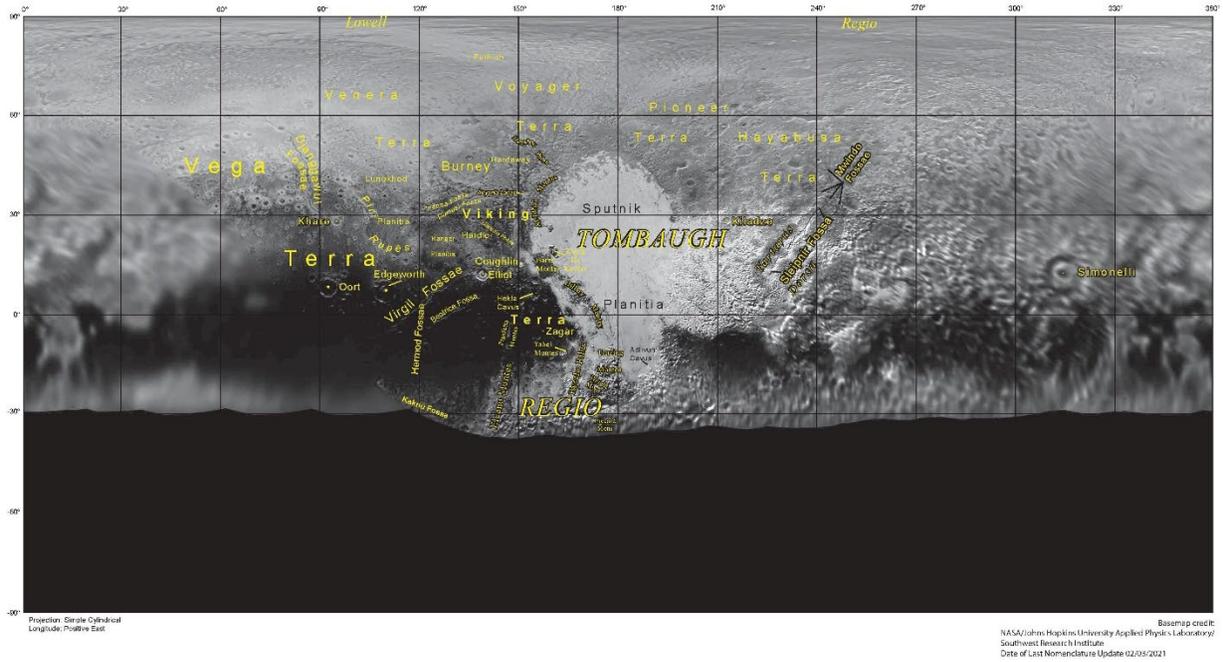

**Fig. A4:** Pluto Nomenclature - the map showing the names of major provinces, geologic units, and features on Pluto. Credit: NASA/Johns Hopkins University Applied Physics Laboratory (JHU-APL)/ Southwest Research Institute (SwRI).

**Section A.6.** The boxplot of the probability of each corresponding 8 gaussian components.

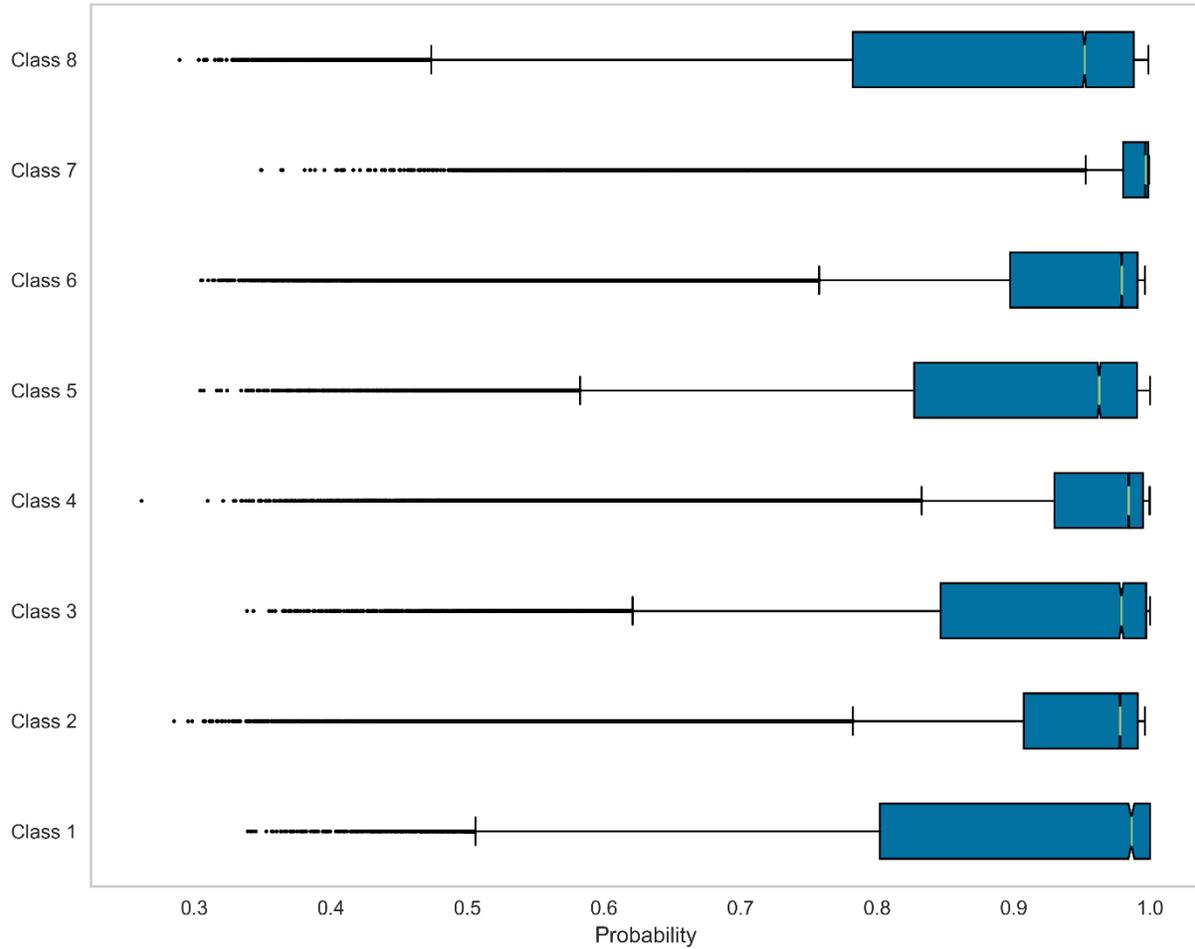

**Fig. A5:** The boxplot of the probabilities extracted from the pixels of corresponding classes (gaussian components). The median is shown by the line that divides the box into two parts while the left and right lines of the box represent the lower quartile (25%) and upper quartile (75%), respectively. The plot shows that most of the surface units are dominated by pixels with higher probability values (close to 1) – indicating the dominance of mostly pure pixels at each unit.

**Section A.7.** The mean ± 1σ standard deviation I/F spectra of subunits for #6 and #3.

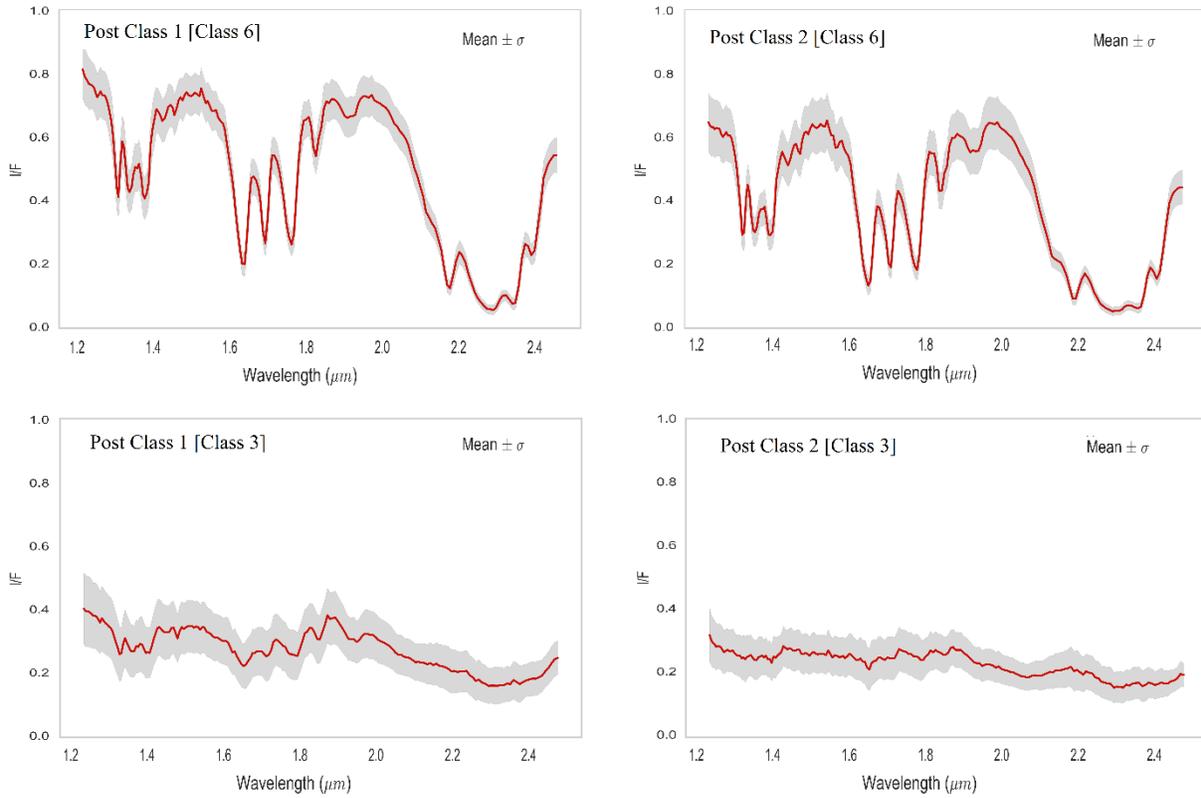

**Fig. A6:** The mean (red line) ± 1σ standard deviation (gray shade) I/F spectra of each subunit for #6 (upper row) and #3 (lower row). The standard deviation of the I/F spectra shows varying degrees of closeness to the mean spectra at different wavelengths.

**Section A.8.** The histogram of the probabilities of corresponding post-classes of C3 and C6.

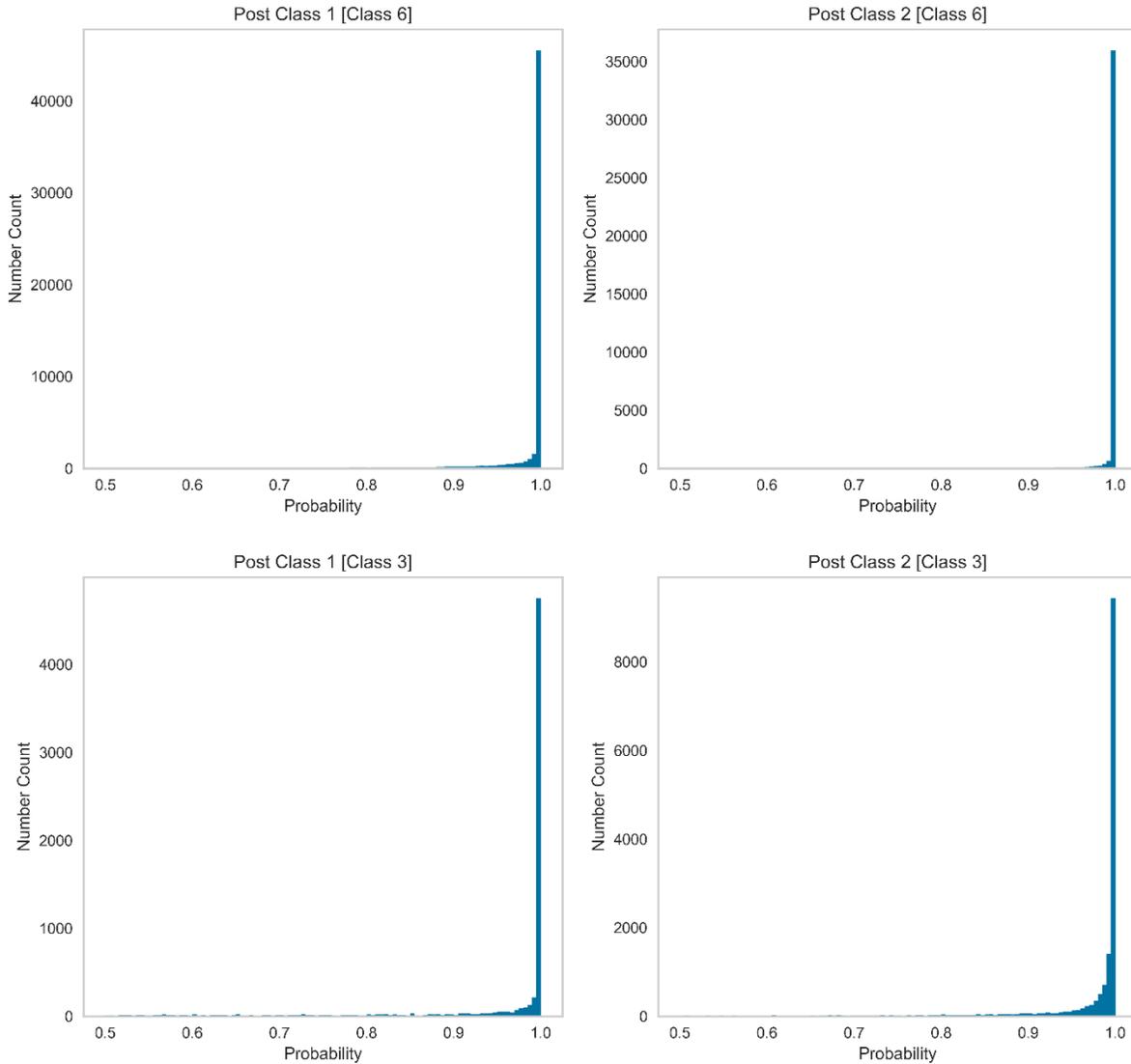

**Fig. A7:** The histogram of the probabilities extracted from the pixels corresponding to each post class of C3 and C6 unit. The plot shows that the surface units are dominated by pixels with higher probability values (close to 1) – indicating the dominance of mostly pure pixels at each unit.